\documentclass{elsart3-1}

\usepackage{amssymb}

\usepackage[english,francais]{babel}
\usepackage[T1]{fontenc}
\usepackage{bm}
\usepackage{graphicx}
\usepackage{tabularx}
\usepackage[dvips]{color}
\usepackage{subfigure}
\usepackage{ulem}

\newtheorem{e-proposition}[theorem]{Proposition}

\newtheorem{e-definition}[theorem]{Definition\rm}

\renewcommand{\theequation}{\arabic{equation}}
\def\og{\leavevmode\raise.3ex\hbox{$\scriptscriptstyle\langle\!\langle$~}}
\def\fg{\leavevmode\raise.3ex\hbox{~$\!\scriptscriptstyle\,\rangle\!\rangle$}}
\def\be{\begin{equation}}
\def\ee{\end{equation}}
\def\ba{\begin{array}}
\def\ea{\end{array}}
\def\bea{\begin{eqnarray}}
\def\eea{\end{eqnarray}}

\def\etal{{\it et al.}}
\def\rmd{{\mathrm d}}

\begin{document}

\begin{frontmatter}


\selectlanguage{english}
\title{Strain localization and anisotropic correlations
in a\\ mesoscopic model of amorphous plasticity}


\selectlanguage{english}
\author[authorlabel1]{Mehdi Talamali,}
\ead{mehdi.talamali@gmail.com}
\author[authorlabel2]{Viljo Pet\"aj\"a,}
\ead{viljo.petaja@gmail.com}
\author[authorlabel1]{Damien Vandembroucq,}
\ead{damien.vandembroucq@espci.fr}
\author[authorlabel3]{St\'ephane Roux}
\ead{stephane.roux@lmt.ens-cachan.fr}

\address[authorlabel1]{Laboratoire PMMH, CNRS-UMR 7636/ESPCI/UPMC/Univ. Paris 7 Diderot\\
10 rue Vauquelin, 75231 Paris cedex 05, France}
\address[authorlabel2]{Laboratoire SVI,  CNRS-UMR 125/Saint-Gobain\\
39 quai Lucien Lefranc, 93303 Aubervilliers cedex, France}
\address[authorlabel3]{LMT-Cachan, ENS de
Cachan/CNRS-UMR 8535/UPMC/PRES UniverSud Paris\\
61 Avenue du Pr\'esident Wilson, 94235 Cachan cedex, France}



\begin{abstract}
A mesoscopic model for shear plasticity of amorphous materials in two dimensions
is introduced, and studied through numerical simulations in order to elucidate
the macroscopic (large scale) mechanical behavior. Plastic deformation is
assumed to occur through a series of local reorganizations. Using a
discretization of the mechanical fields on a discrete lattice, local
reorganizations are modeled as local slip events. Local yield stresses are
randomly distributed in space and invariant in time. Each plastic slip event
induces a Eshelby-like long-ranged elastic stress redistribution. Focusing on
quasistatic loadings and zero-temperature limit, extremal dynamics allows for
recovering many of the complex features of amorphous plasticity observed
experimentally and in numerical atomistic simulations in the quasi-static
regime. In particular, a quantitative picture of localization, and of the
anisotropic strain correlation both in the initial transient regime, and in the
steady state are provided.

\vskip 0.5\baselineskip

\selectlanguage{francais}

\vskip 0.5\baselineskip \noindent {Nous pr\'{e}sentons un mod\`{e}le
  num\'{e}rique \`{a} l'\'{e}chelle mesoscopique de la plasticit\'{e}
  des amorphes. La d\'{e}formation plastique est suppos\'{e}e
  r\'{e}sulter de s\'{e}ries de r\'{e}arrangements locaux de la
  structure.  La discr\'{e}tisation sur r\'{e}seau des champs
  m\'{e}caniques permet d'assimiler ces r\'{e}arrangements \`{a} de
  simples glissements locaux. Les seuils de contraintes associ\'{e}s
  \`{a} ces \'{e}v\'{e}nements sont al\'{e}atoires et ind\'{e}pendants
  dans l'espace comme dans le temps. Chaque \'{e}v\'{e}nement
  plastique induit un champ de contrainte \'{e}lastique \`{a} longue
  port\'{e}e (type Eshelby). S'int\'{e}ressant au comportement
  quasistatique et \`{a} la limite d'une temp\'{e}rature nulle,
  l'utilisation d'une dynamique extr\'{e}male permet de rendre compte
  de nombreux ph\'{e}nom\`{e}nes observ\'{e}s exp\'{e}rimentalement et
  num\'{e}riquement dans le r\'{e}gime quasi-statique. En particulier
  nous pr\'{e}sentons des r\'{e}sultats quantitatifs sur le
  comportement de localisation et l'anisotropie des corr\'{e}lations
  de la d\'{e}formation plastique.}

\vskip 0.5\baselineskip
\noindent{\small{\it Mots-cl\'es~:} Mot-cl\'e1~de~la~liste~; Mot-cl\'e2~;
Mot-cl\'e3}}
\end{abstract}
\end{frontmatter}


\selectlanguage{english}
\section{Introduction}
\label{intro} The prominent mechanical property of glasses is their brittleness.
Nevertheless, the question of their possible plasticity has been
discussed very early\cite{Taylor-Nat49,Marsh-64}.  Limited to small
scale or confined geometries (due to their brittleness), the plastic
behavior of glasses is quite different from its crystalline
counterpart. In particular, a notable pressure-dependence is often
encountered and, in addition to shear flow, it is common to observe
permanent densification\cite{Ernsberger-JACS68,Peter-JNCS70}.  Despite
this early interest for the plastic behavior of oxide glasses and the
primary importance of this phenomenon for mechanical contact
properties, the subject has long remained poorly explored. The recent
development of metallic glasses has given a new momentum to this
subject~\cite{Schuh-ActaMat07}. Indeed, the mechanical strength of
metallic glasses appears to be limited by shear-banding induced
failure phenomena.

One of the most important issues that constitute a barrier in the
development of those materials is the proper mastering of size
effects.  Indeed, if the yield limit appears to be extremely high as
compared to crystalline materials with the same chemical composition,
yet failure appears to be quite brutal for large scale specimens, akin
to brittle fracture.  Amazingly, a closer analysis show that these
materials may withstand a large number of macroscopic slips over
different shear planes prior to failure.   The proper understanding of
the initiation of inhomogeneous plastic strain is thus essential to
enhance the mechanical properties of these materials.

{In terms of theoretical modeling, a major difficulty long
  lied in the absence for amorphous materials of any obvious
  microscopic alternative concept to dislocations for crystalline
  plasticity. The understanding that plastic deformation resulted from
  a series of local structural
  reorganizations~\cite{Argon-ActaMet79,FalkLanger-PRE98}, commonly
  referred to as ``shear transformation'', (ST), was a major
  theoretical breakthrough introduced by Argon. This new paradigm,
  together with the fast
  development of computing facilities has opened the way to direct
  simulations of shear plasticity of glasses at atomistic scale (such
  as Molecular Dynamics) and mesoscopic scale (intermediate between
  the size of a structural reorganization and the continuum scale).}

The phenomenology of amorphous plasticity that emerged frome these
numerical studies to be discussed in greater detail below (see
also Ref. \cite{RTV-MSMSE11} for a review) is rather complex and
questions the traditional view of plastic flow as smooth and
regular. In particular phenomena such as intermittence, avalanche or
quasi-localisation are shown to occur over a wide range of time- and
length-scales. The observation of such scaling behaviors naturally
calls for a description which necessarily includes fluctuations and
not simply averages, even at macroscopic scales.

{In the framework of the complex phenomenology of amorphous plasticity,
our ambition is thus the construction of the minimal model that
reproduces a number of macroscopic features.  By minimal, we mean that
the least number of free parameters and computational complexity is
our guide.  For this we will have to sort between important features
which survive at a macroscopic scale, and others (which may indeed
exists) which will be disregarded if they do not impact the large
scale behavior. Following an early work by Baret
\etal~\cite{BVR-PRL02}, our model will be based on the competition
between the respective effects of the structural disorder of the
amorphous matter and of the elastic coupling induced by the reaction
of the elastic bulk to local reorganizations\cite{Eshelby57}.}

In this paper, a particular emphasis is put on the geometrical
features of the plastic strain field.  We first detail the ambition
and the limits of our modeling approach in section~\ref{sec:intro_model}. The
model is constructed in Section~\ref{sec:model_def}, the stress-strain
relation is discussed in Section~\ref{sec:stress-strain}, the
localization behavior is discussed in Section~\ref{sec:localization}.
Section~\ref{sec:strain_correlation} gives a quantitative
characterization of the anisotropic strain correlation. A final
discussion, Section~\ref{sec:conclusion}, concludes this article.

\section{Amorphous plasticity: from microscopic scale to macroscopic scale}\label{sec:intro_model}

In this section we briefly discuss the complex phenomenology of
amorphous plasticity that has emerged from the large body of numerical
simulations performed in recent years and we introduce our modeling
strategy.

\subsection{Molecular Dynamics}

In view of the obvious numerical limitations of atomistic simulations, most of
the recently simulated systems consisted of 2D Lennard-Jones model glasses or
close
variants~\cite{FalkLanger-PRE98,Tanguy-PRB02,Varnik-PRL03,Maloney-PRL04a,Maloney-PRL04b,Tanguy-PRB04,Falk-PRL05,Tanguy-PRB05,Maloney-PRE06,Tanguy-EPJE06,Falk-PRL07,Maloney-JPCM08,Tsamados-EPJE08,Tsamados-PRE09,Maloney-PRL09,Lemaitre-PRL09,Lerner-PRE09b}
which already capture most of the difficulties of the problem, and yet allowed
to consider large system sizes.  A few three dimensional
works~\cite{Varnik-JCP04,Falk-PRB06} indeed confirmed that little new physics
emerged from the third dimension. Some simulations even made use of more
``realistic'' potentials for metallic glasses~\cite{Schuh-ActaMat03}, amorphous
silicon,~\cite{Argon-PRL04,Argon-PRB05a,Argon-PRB05b,Talati-EPL09}, vitreous
polymers~\cite{Rottler-PRE01,Rottler-PRE03} or silica
glass~\cite{Tanguy-PRL06,RVTBR-PRL09} reaching similar conclusions.

Although a wide consensus exists on the scenario of ST {initially
proposed by Argon~\cite{Argon-ActaMet79}} as validated by the numerous numerical
studies~\cite{Tanguy-PRB02,Argon-PRL04,Maloney-PRL04a,Maloney-PRL04b,Argon-PRB05a,Argon-PRB05b,Maloney-PRE06,Tsamados-EPJE08,Tsamados-PRE09,Talati-EPL09,Delogu-PRL08,Rodney-PRL09},
the quantitative characterization of these local reorganizations as well as
their links to the local structure remains quite elusive.

Beyond the question of the microscopic mechanisms, a special attention
has been devoted to the complex character of the plastic behavior of
glasses.  At a microscopic scale, the ``non-affine'' corrections to
the elastic behavior have been thoroughly
discussed~\cite{Tanguy-PRB02,Maloney-PRL04a,Maloney-PRL04b,Tanguy-PRB04,Tanguy-PRB05,Maloney-PRE06,Tanguy-PRL06,Goldenberg-EPL07,Tsamados-EPJE08,Tsamados-PRE09}
as well as their possible consequences on the initiation of plastic
reorganizations~\cite{Tsamados-PRE09}. Strain localization has
received a particular
interest~\cite{Varnik-PRL03,Falk-PRL05,Falk-PRB06,Maloney-PRL09,Lerner-PRE09a}
as well as avalanche statistics~\cite{Maloney-PRL04a,Lemaitre-PRL09};
akin to other non-linear mechanical behaviors in random media
(e.g. fracture), plastic deformation is not continuous but proceeds by
successive bursts.

\subsection{Meso-models}

In parallel to the above discussed atomistic simulations, several works have
developed numerical models at a mesoscopic scale, {\it i.e.}
 intermediate between the scale of an individual reorganization and the
macroscopic scale where continuum mechanics gives a faithful description.
Assuming that plastic deformation resulted from a series of reorganizations,
Argon and Bulatov early proposed a discrete model of amorphous plasticity and
studied its behavior at different
temperatures~\cite{BulatovArgon94a,BulatovArgon94b,BulatovArgon94c}. Homer and
Schuh~\cite{Schuh-ActaMat09} very recently proposed an extension of this model
to follow the dynamics of plastic deformation of amorphous metals under shear
using a kinetic Monte-Carlo algorithm.

Along similar ideas, Baret \etal~\cite{BVR-PRL02} developed a model of amorphous
plasticity at zero temperature in antiplane geometry with two main ingredients:
a structural disorder and a long-range elastic interaction to account for the
stress redistribution due to local plastic reorganizations. This model, which
exhibits critical properties, can also be considered as belonging to the same
vein as other statistical modeling for earthquakes~\cite{Chen-PRA91,BZR-JGR93}
or elastic line depinning~\cite{Narayan-PRB93,Kardar-PR98}.  In a close spirit,
Picard  {\it et al} discussed complex spatio-temporal behavior of a yield stress
fluid~\cite{Picard-PRE05}. In this model, the authors considered a quadrupolar
elastic interaction but no other source of disorder than the one caused by the
relaxation dynamics. Following similar ideas, Bocquet {\it et al} developed an
analytic model for the elastoplastic dynamics of a jammed material taking into
account the elastic stress redistribution\cite{Bocquet-PRL09}. Lema\^{\i}tre and
Caroli\cite{Lemaitre-preprint06} recently discussed the avalanche behavior
within a mean field model. In particular, they proposed to link ideas of
effective temperature discussed by Sollich \etal~\cite{Sollich-PRL97} to
describe the rheology of soft glasses and the internal stress fluctuations by
varying the distribution of their elastic interaction.  Jagla~\cite{Jagla-PRE07}
discussed the effect of a local relaxation processes on the localization
behavior. Sekimoto\cite{Sekimoto-PRL05} discussed the internal stress as
encoding past mechanical treatments. In a more traditional dislocation
plasticity spirit, Zaiser and Moretti\cite{Zaiser-JSM05} discussed the avalanche
behavior in a mesoscopic model very close to the above discussed ones. More
recently Dahmen {\it et al}\cite{Dahmen-PRL09} proposed a variant of the mean
field Ben-Zion--Rice earthquake model\cite{BZR-JGR93} incorporating a systematic
weakening or hardening effect.


\subsection{Macroscopic scale: softening vs hardening}

 Some of these studies have been dedicated to the
characterization of the plastic behavior within a continuum mechanics
framework~\cite{Rottler-PRE01,Rottler-PRE03,Schuh-ActaMat03}.

The large scale behavior is observed to approach a so-called elastic-perfectly
plastic constitutive law, where the macroscopic stress becomes independent from
the macroscopic strain.  However, the latter law lies at the edge of mechanical
stability. Would stress increases with strain (``hardening'' behavior) that the
macroscopic strain field would tend to be homogeneous {and
deterministic} at large scales (under a uniform stress) amenable to
homogenization\cite{Zaoui-JEMT84,Zaoui-JEM02}. In contrast, a ``softening''
behavior (where stress decreases with strain) induces at large scales
localization modes where strain concentrates on shear bands. {In the
latter case, homogenization is inappropriate, and the
  macroscopic behavior results from microscopic details (initiation
  and growth of the shear band) without much
  hope for a picture which would uniformly hold for a large class of
  amorphous materials.}

  The perfectly plastic behavior lies at the
borderline between these two classes of behavior. Hence, one cannot state any
general results about the homogeneity of the strain field or the fate of a
localization mode. {It is necessary to resort to the scaling features
of a meso scale modeling as it will determine the answer to such a question.
Interestingly enough, this perfect plasticity character will justify the fact
that fluctuations will survive up to the macroscopic limit, without scale
cut-off.  Hence, in contrast to many other classes of constitutive law, an
enrichment capturing fluctuations at all scales is a need.}

\subsection{Modeling strategy}

The vanishing of the tangent {macroscopic} stiffness is thus the key
feature that allows for the relevance of more microscopic details than in
hardening behaviors. However, not all details matter, and sorting out the
important ingredients needed to enrich an effective macroscopic description is
an important issue which has not been addressed as such in the abundant above
mentioned literature.

In conclusion, a clear view of up-scaling is still missing. This question is
limiting inasmuch as the identification of a proper continuum description
framework is not settled ({\it e.g.} is the plastic strain field uniformly
distributed or rather localized in the thermodynamic (large size) limit?) and
hence the question of what microscopic features survive at a macroscopic scale
(and should be extracted from extensive microscopic studies) is not elucidated
yet. The present work aims at addressing the question of the asymptotic
homogeneity or heterogeneity of the plastic strain over various space and time
scales.  In order to approach this problem, we will consider a large scale
modeling and argue for the relevance at large scales of the elementary
mechanisms in line with microscopic or mesoscopic modeling.

We will consider a large scale modeling and argue for the relevance at
large scales of the elementary mechanisms in line with microscopic or
mesoscopic modeling. Our model will be seen to reproduce the early
work by Baret {\it et al}~\cite{BVR-PRL02} in the context of
anti-plane loading.  The latter case corresponded to a parallel
displacement field along say the $z$-axis, whose amplitude depends
only on the transverse coordinates $U_z(x,y)$.  We will here rather
focus here on a different symmetry, e.g.  plane strain, where $U_z=0$,
and the in-plane displacement depends only on $x$ and $y$.  Because
the modeling is proposed based on large scale asymptotics, we will
retain only the most salient effects which have a chance to survive at
a large scale and hence, our objective is to propose the minimal level
of complexity. Nevertheless, a detailed comparison of scaling features
with microscopic modeling such as that of Maloney and
Robbins\cite{Maloney-PRL09} can be performed and will be shown to be
extremely close. {Alternatively, such a mesoscopic model
  should not be used to address questions pertaining to the
  microscopic scale.  Therefore, without questioning their validity, a
  number of microscopic properties reported in the abundant literature
  will not be retained if we do not anticipate a macroscopic
  manifestation of these observations.  These choices should not be
  interpreted as a negation of a more complex microscopic reality than
  the one we propose here.  On the contrary, our goal is to identify
  the most elementary model, based on some accepted observations,
  which is amenable to reproduce the large scale features of amorphous
  media plasticity.  Part of this motivation is to be able to
  investigate asymptotic scaling features.  For instance, the question
  of the asymptotic nature of the plastic strain distribution
  (homogeneously distributed as for hardening behaviors, localized as
  for softening behaviors, or a more subtle intermediate regime to be
  explicited) is at the very heart of our approach.}

\section{Definition of the model}\label{sec:model_def}

The studied model is here constructed with the objective of capturing the {\em
large scale behavior} of amorphous media.  In order to limit the computational
cost of studying the model, only the two dimensional case of plane strain will
be considered.  It is recalled that similar behaviors where observed in
molecular dynamics simulations in two and three dimensions, and hence this
restriction is not limiting.  As argued in the introduction, the macroscopic
loading may induce changes in volume which are coupled with the activation of
plastic flow.  However, for extended shear, no macroscopic dilation or shrinkage
can be expected and hence the dominant macroscopic plastic strain will be purely
deviatoric. {Note that a macroscopic volume change may occur in the
transient early deformation regime, but is not accounted for in our model.
Because only the transient regime would be affected, the scaling analysis of the
steady plastic flow should not be affected.} Hence for simplicity pure shear
boundary conditions are considered, with fixed principal axes in order to reach
large strains. More precisely the mean total strain field is controlled as
    \be
    \langle\bm\varepsilon\rangle=\varepsilon\left(\ba{cc} 1 &0\\ 0 &-1\ea\right)
    \ee
In the following a discretization will have to be introduced to study
numerically the model. A square lattice with bi-periodic boundary conditions
will be used to avoid a systematic spatial bias induced by rigid boundaries.

The microscopic structure of amorphous media is intrinsically heterogeneous, and
hence, even in the absence of plastic reorganization, the elastic behavior of
the material is heterogeneous.  However, upon coarse-graining, this
heterogeneity will vanish giving rise to an homogeneous behavior for which
elastic constants can be derived from classical homogenization techniques. Thus,
the elastic regime will be described as homogeneous. Moreover, the amorphous
medium will be assumed to be isotropic. Although this has been questioned on the
basis of molecular dynamics simulation for amorphous silica~\cite{RVTBR-PRL09},
such an anisotropy is weak and does not affect the main features (and in
particular scaling properties) of the model at large scales.  Since no volume
change will be considered, only one elastic constant, the shear modulus, $\mu$,
has to be introduced, which {in a dimensionless model can be set to
$2\mu=1$}.  {Note that this statement does not contradict the existence
of elastic heterogeneities, and their role in the triggering of ST,  but it
merely says that those heterogeneities do not survive at a large scale.  Thus we
may simply resort to a deterministic homogeneous description at a large scale.
Hence whenever we refer to a local stress, it will refer to that of the
homogenized picture.  The actual local stress would be related to the
homogeneous one with a localization operator which is not introduced explicitly
because it is not needed in the sequel. }

The shear transformations are the only source of non-linearity in the model, and
hence at any stage, the medium can be unloaded and reloaded by the mere
superposition of a uniform shear stress field (with a fixed orientation). Hence,
the tensor stress field can be characterized by its sole deviatoric norm (or
equivalent von Mises stress).  Scalar (equivalent) stress $\sigma\equiv
\sigma_{xx}-\sigma_{yy}$ and strain $\varepsilon\equiv
\varepsilon_{xx}-\varepsilon_{yy}$ will be used (although we use a plane strain
assumption, the conditions specified below will guaranty that plane stress also
hold). The latter scalar stress component will be denoted as $\sigma$, and
called ``stress'' for simplicity. Similarly a single scalar strain component
will matter, and will be denoted as the strain, $\varepsilon$, in the sequel.
With those notations the incremental elastic law reduces to the elementary form
$\Delta\sigma=\Delta\varepsilon$.

As above discussed, in amorphous materials
  such as glasses, pastes or foams, a common assumption consists in
  describing the macroscopic deformation as deriving from a succession
  of localized reorganizations at some microscopic
  scale\cite{FalkLanger-PRE98}. These regions can be seen as the
  coarse grained Shear Transition Zones (STZ)~\cite{FalkLanger-PRE98}
  whose sizes reach at most a few tens of atoms~\cite{Tanguy-EPJE06}
  in molecular dynamics simulations. The details of such local
  rearrangements depend obviously on the precise structure of the
  material under study. They involve local change in the topology of
  atoms, grains or foam cells, and significant non-linearities either
  of geometrical or constitutive origin take place.  The important
  feature is that these reorganizations are local is space.  We
  discuss below the criterion to initiate such an event
  (Sect.~\ref{ssec:onset}) and its effect on the stress field
  (Sect.~\ref{ssec:effect}).

\subsection{Onset of local plastic transformation}\label{ssec:onset}

In order to describe at which stage of loading such a shear transformation will
take place, a criterion has to be proposed.  It is natural in the present
framework to characterize locally the onset of a transformation through the
local stress.  {Here local refers to the homogenized description}. In
addition, it is also at the discretization scale.  However, our discretization
is itself at a mesoscopic scale not to be confused with the microscopic one. Let
us only note here that the full characterization of the local reorganization
occurring under shear in amorphous materials is still an ongoing issue. In
particular the existence for amorphous materials of a local stress threshold at
atomic scale has recently been debated~\cite{Tsamados-EPJE08,Rottler-PRE10}.
While the definition of a satisfactory criterion can be discussed at atomic
scale, there is no doubt however that under sufficient coarse graining, one
should recover a criterion based upon the {\em homogenized} stress
tensor {whatever the localization tensor}. Irrespectively of the
precise characterization of local instabilities at atomic scale, we assume in
the present mesoscale model, that the coarse-graining is performed at a large
enough length scale to allow us to safely use local yield stress criteria. The
criterion for yielding characterizes the local configuration of atoms, and hence
will display some variability.  A local yield threshold for each discrete site
$\bm x$ as $\sigma_{\gamma}(\bm x)$ is introduced, and will be treated as a
random variable in the sequel. For all sites, the same statistical distribution
will be used, chosen for simplicity as a uniform distribution over the interval
$[0,1]$. Other distributions could be considered but are not expected to alter
the generic behavior of the model beyond a few mesh sizes (and this has been
checked numerically).

It is to be noted that since no stress scale has been introduced in the model so
far, rescaling the distribution to $[0,\varsigma_1]$ will only multiply all
stresses by the same factor $\varsigma_1$.  Similarly, the interval may be
translated to $[\varsigma_0,\varsigma_0+\varsigma_1]$, by adding a constant
$\varsigma_0$ to all stresses. Thus the linearity of the model can be exploited
to match any yield stress interval. More importantly, other distributions, such
as a Gaussian, behave similarly, and at a large enough scale, the local yield
distribution can no longer be read from microscopic observables, as a
consequence of universality.

\subsection{Effect of a local transformation} \label{ssec:effect}

Beyond the confined region where the rearrangement takes place, a stress
perturbation has to be accommodated by elastic strain throughout the material.
This perturbation induces internal elastic stresses $\sigma_{el}$ which vanish
with the distance, as dictated by linear elasticity (of a homogeneous medium as
above argued). The general solution of elastic field induced by a localized
perturbation can be written easily as discrete infinite series (multipolar
expansion~\cite{TPRV-PRE08}). An important feature of those modes is the absence
of length scale. Therefore, these stress influence functions can be extrapolated
down to a point-like singularity.  As such, they are genuine Green functions
attached to the elementary rearrangements. The constraint of not imposing
externally any force or torque selects fields which decreases with the distance
to the transformation zone as $r^{-n}$ where $n\ge D$ where $D$ is the space
dimension. The leading perturbation term is hence given only with the $n=D$
terms.  Three such modes exist in two dimensions, $D=2$, which coincide with the
external problem of a plastic strain imposed uniformly in a circular inclusion,
the so-called Eshelby problem\cite{Eshelby57}.  One is a pure dilation which
corresponds to the expansion or shrinkage of the transformation zone. The two
others are pure shear modes with different principal axes.  Because the
macroscopic loading is a pure shear, and large plastic strains are considered,
the dilation mode will vanish on average.  Thus, any change of volume will be
discarded.  Similarly, on average, the orientation of the deviatoric mode will
be aligned with the macroscopic loading, and fluctuations around this
orientation will be neglected.  Consequently, only one mode will be retained to
describe the elastic effect of a localized transformation. {Let us
underline that neglecting the two other Eshelby modes is a simplifying
assumption.  Introducing those modes would require a full tensorial framework
for stress and strains, and hence a much more complex description for triggering
an elementary plastic event.}  Mesoscopic scale models such as
\cite{BVR-PRL02,Picard-PRE05} thus consist in studying the effect of these
dominant terms of the elastic interaction on the dynamics of localized plastic
events, independently of the microscopic details. It is stressed that there is
no upper limit on the discretization scale, since the considered external
Eshelby mode does not involve any specific length scale. However, if the
discretization scale changes, the corresponding statistical features of the
local rearrangements have to be coarsened at the appropriate scale and hence the
microscopic disorder has to be adjusted to match a specific material. This
question will not be addressed in the present study.

An elementary local transformation zone occurring at point $\bm x_0$ is thus
described by a stress field $\eta G(\bm x-\bm x_0)$ whose scalar amplitude
$\eta$ is the only relevant feature.  Note that the translational invariance
which is natural in an infinite medium still holds for the chosen bi-periodic
boundary conditions in a finite size system.  The precise form of $G$ is
discussed below in Sect.~\ref{ssec:Green}. Reverting to the original Eshelby
problem, this amplitude can be interpreted as the product of a uniform plastic
strain within a circular inclusion times its volume (or area in our
two-dimensional framework). But the key point, is that the external stress field
is independent of the specific shape of the inclusion and details of the
re-arrangement --- such dependencies only appear in higher multipolar orders.
Nor does it make sense to distinguish separately volume and local plastic
strain. In the sequel, conventionally the volume of the STZ can thus be chosen
to be equal to the volume of the discretization scale, namely unity, and hence
the amplitude $\eta$ is a measure of the equivalent local plastic strain,
denoted as ``local slip'' in the following.

An elementary local transformation of amplitude $\eta$ occurring at point $\bm
x_0$ induces a local increase in the plastic strain field
    \be
    \Delta\varepsilon_p(\bm x)=\eta \delta(\bm x-\bm x_0)
    \ee
where $\delta$ is the Dirac distribution.  The residual stress field
$\sigma_{el}$ is correspondingly modified by
    \be
    \Delta\sigma_{el}(\bm x)=\eta G(\bm x-\bm x_0)
    \ee
This residual stress field itself encodes a memory of the set of transformations
experienced by the medium, and long-range spatial correlation will naturally
accumulate in the stress field.  Thus at any stage of loading the local stress
consists of a homogeneous macroscopic stress, $\Sigma$, added to the residual
stress field,
    \be
    \sigma(\bm x)=\Sigma+\sigma_{el}(\bm x)
    \ee
As local transformations can occur anywhere in space, each point $\bm x$ is
attributed a threshold stress $\sigma_{\gamma}(\bm x)$.  Starting from rest, at
each instant of time, we deduce the level of external loading needed to trigger
a local transformation as
    \be\label{eq:criticalloading}
    \Sigma_c=\min_{\bm x}\left[\sigma_{\gamma}(\bm x)-\sigma_{el}(\bm x)\right]
    \ee

The intrinsic local disorder of amorphous media implies that the slip amplitudes
$\eta$ are randomly distributed. Moreover, because our description is sought at
a large scale no spatial nor temporal correlations will be considered. To
account for the local reorganizations, it thus suffices to give the probability
distribution function of their (scalar) amplitude.  The only physical constraint
is that they should be bounded.  Similarly to the yield thresholds, the slip
value $\eta$ is drawn randomly from the uniform distribution, $[0, d]$ if not
otherwise stated. $d$ is a parameter of the model whose role will be discussed
below.

\subsection{Extremal dynamics}

Quasi-static driving conditions are considered.  Thus stress redistribution is
considered to occur instantaneously (no viscous effects are considered).
However, one should finally specify the way the system is driven.  A constant
external stress is inappropriate as one expects the stress strain law to tend to
a constant plateau stress.  An imposed total strain is more relevant, but it has
the drawback of possibly triggering multiple transformation zones.  We have
chosen the so-called ``extremal dynamics'', which consists of adjusting at each
instant of time the external loading $\Sigma$ so that one and only one
transformation zone $\bm x^*(t)$ is activated at each time $t$ according to
Eq.~\ref{eq:criticalloading}.  This may involve a reduction in the applied
macroscopic stress and strain.  A unique temporal sequence of event is selected,
from which it is easy to reconstruct any other driving mode assuming the same
sequence of events.  In this way, ``avalanches'' can be defined without
ambiguity.  Note however, that ``time'', $t$, is used here as a simple way of
counting and ordering events.  On average, time is simply proportional to the
total plastic strain imposed on the system,
$\langle\varepsilon_p\rangle=tdL^{-2}/2$. The plastic strain field is thus
simply
    \be
    \varepsilon_p(\bm x,t)=\sum_1^t \eta(t)\delta(\bm x-\bm x^*(t))
    \ee
This specific driving mode is the one that would result from an over-dampened
viscous dynamics and infinitesimal strain rate driving.

\subsection{Stress redistribution function}\label{ssec:Green}

\begin{figure}[htbp]
\subfigure[]{
  \begin{minipage}[h]{0.3\textwidth}
   \centering
  \includegraphics[width=50mm]{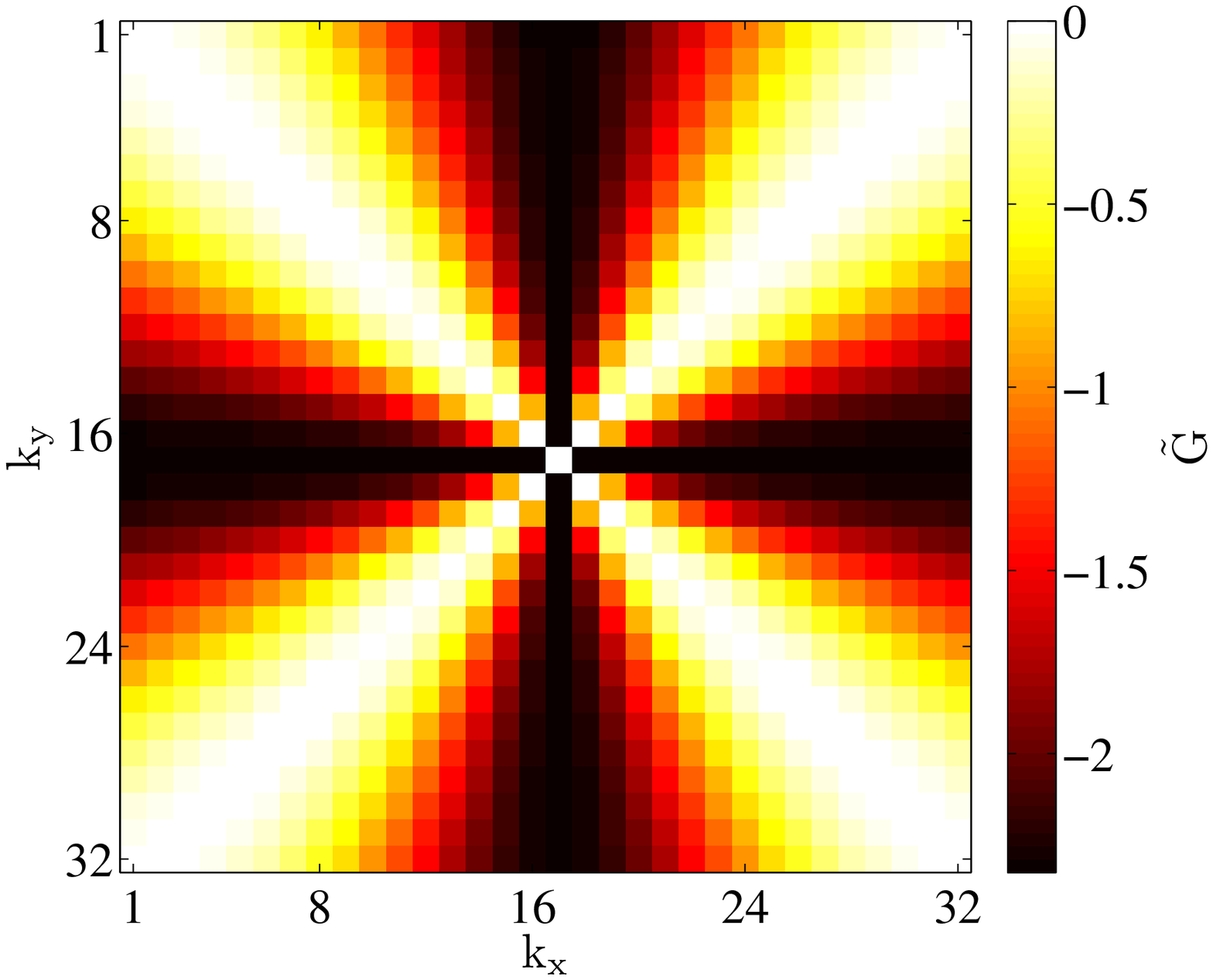}\\
  \end{minipage}
  \label{Fourier}}
 \subfigure[]{
  \begin{minipage}[h]{0.3\textwidth}
   \centering
   \includegraphics[width=50mm]{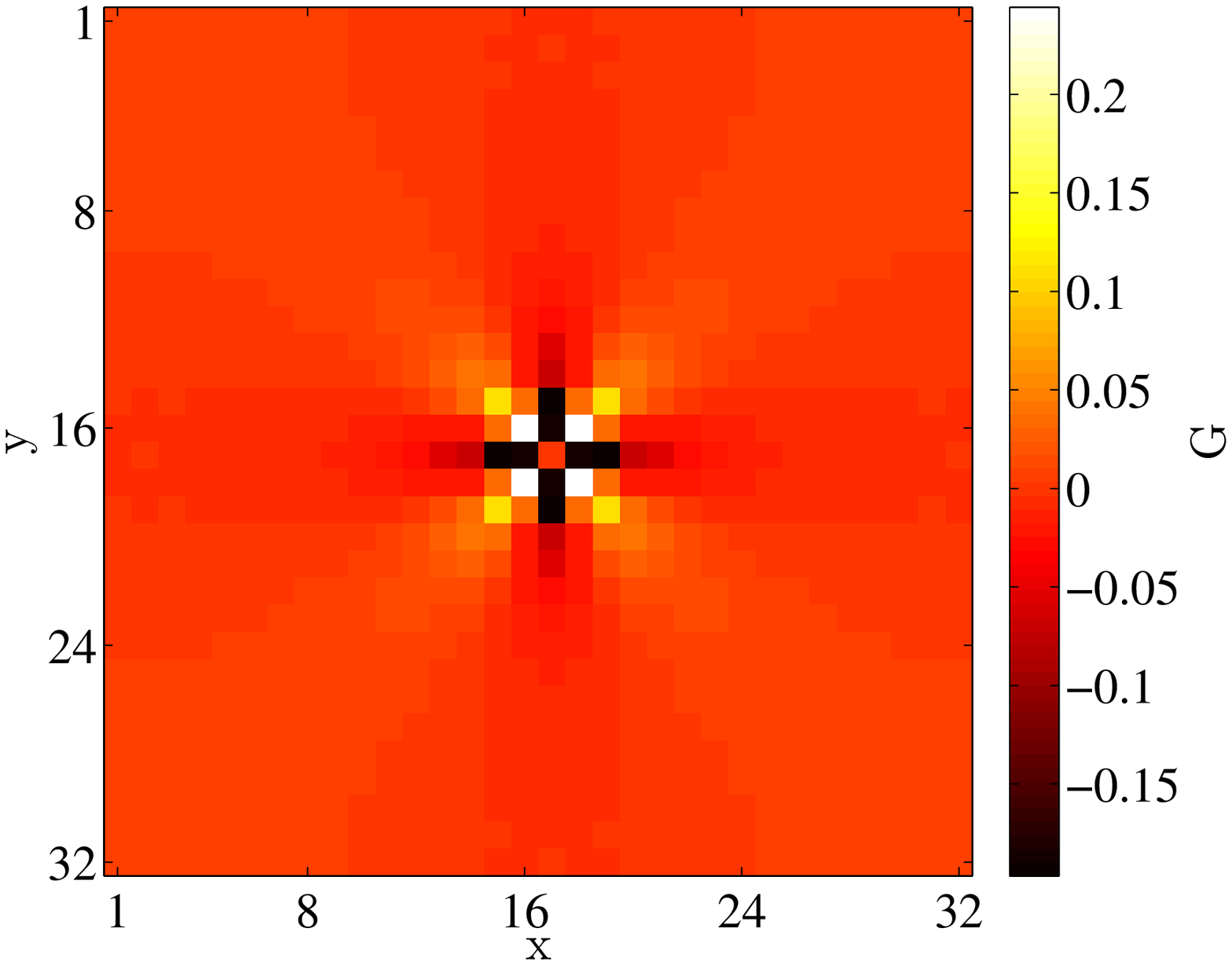}
  \end{minipage}
  \label{direct}}
 \subfigure[]{
  \begin{minipage}[h]{0.3\textwidth}
   \centering
   \includegraphics[width=50mm]{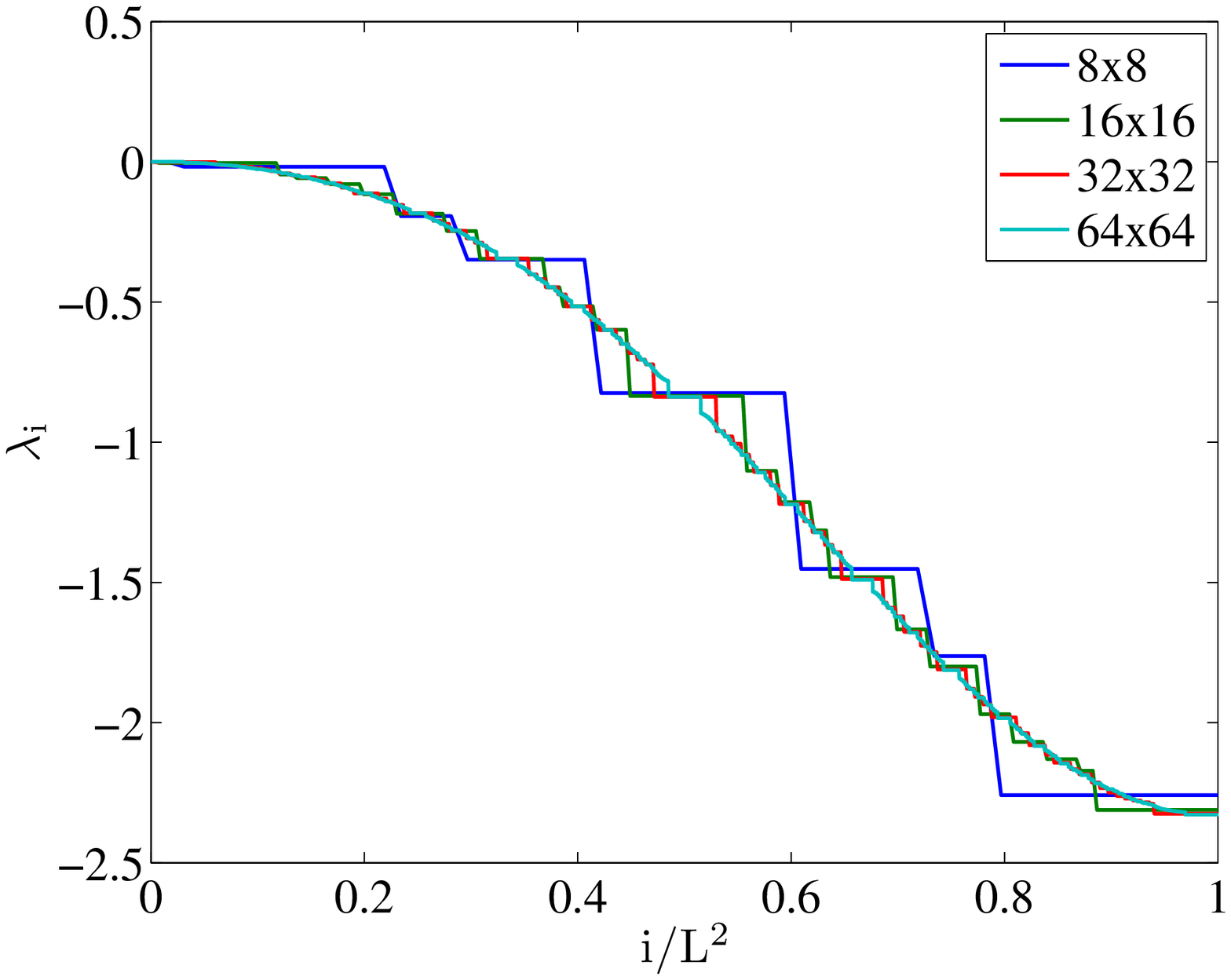}
  \end{minipage}
  \label{spectrum}}
  \caption{Map of the quadrupolar elastic interaction used in the
    model. The discretization is performed in the Fourier space $\tilde{G}
    (k,\omega) = cos(4\omega)-1$, where $k$ and $\omega$ are the polar
    coordinates in the Fourier space ($k_x,k_y)$ (a) and a subsequent Fourier transform
    gives in the direct space the Green function $G(x,y)$  (b)
    satisfying the bi-periodic boundary conditions of the problem.
    (c) Spectrum of the eigenvalues of the elastic propagator. The
    horizontal tangent at the maximum eigenvalue zero can be associated
    to an abundance of soft modes (with null or close to zero eigenvalues).
    The associated eigen-modes are aligned along directions equal or
    close to $\pm 45^\circ$.}
  \label{Green}
\end{figure}

The Green function $G$ can easily be computed to be $G=\cos(4\theta)/r^2$ in
polar coordinates for an infinite isotropic medium. The quadrupolar elastic
kernel is obviously a key ingredient of the present model. As already discussed
by Picard {\it et al} \cite{Picard-PRE05}, the numerical implementation of this
kernel is rather delicate. In order to deal with the bi-periodic boundary
conditions we chose here to discretize first in the reciprocal space
$\widetilde{G}(k,\omega) = -\cos(4\omega)-1$, where $k$ and $\omega$ are the
polar coordinates in the Fourier space, (Fig.~\ref{Green}a) then to Fourier
transform to get the periodic kernel $G(r,\theta)$ in the direct space
(Fig.~\ref{Green}b).
In the present case, because of the bi-periodic boundary conditions, the elastic
propagator is a circular matrix. The eigenvalues of this (convolution) operator
are directly obtained by Fourier transforming the associated Green function. At
the same time as the elastic propagator we thus obtain the spectrum of its
eigenvalues, as shown in Fig.~\ref{Green}c, without performing any further
computation.

The residual stress field resulting from an arbitrary cumulative plastic strain
is easily written in Fourier space as
    \be
    \widetilde \sigma_{el}=\widetilde{G}.\widetilde{\varepsilon_p}
    \ee
Hence integration over a long interval of time shows that a stationary state can
be reached where $\sigma_{el}$ remains constant, while plastic strain
accumulates over the modes associated with a zero eigenvalue for $G$.  This
shows trivially that uniform shear along lines at $\pm \pi/4$ orientation with
respect to the $x$-axis, but with arbitrary values from line to line meets this
stationary condition.  However, as plasticity develops in our model by discrete
events of random amplitude, it is not possible to fulfill such condition at each
instant of time.  The ``granularity'' of the elementary events imposes a
locality of plastic strain in real space, which in turn is non-local in Fourier
space, and hence, stress fluctuations will occur for which it is essential to
account faithfully for the vicinity of the spectrum of eigenvalues close to 0.
This point is emphasized here because slight variations in the discretization of
the Green function $G$ which may appear as innocuous have a drastic influence in
the long range scaling properties of the model.

\subsection{Slip amplitude distribution}

Let us come back to the meaning of the parameter $d$ which is the only
dimensionless parameter of the model. It quantifies the elementary plastic
strain required to change the conformation of the local shear transformation
zone, at the scale, $a$, of our discrete mesh in such a way that local yield
stress correlation over time can be ignored.  After one elementary local
transformation event, the stress fluctuation for a site neighbor to the slip is
of order $d$. This quantity is to be compared to the amplitude of the yield
stress fluctuations, $\varsigma_1$. Hence the important dimensionless parameter
is $d/\varsigma_1$. In the sequel, the yield stress amplitude $\varsigma_1$ is
conventionally set to unity. Therefore, $d$ remains the only free parameter of
the model. For most of the simulations shown below, a value of $d=0.01$ is
chosen.

In the following we will focus on the relation between the local
plasticity events and the macroscopic plastic flow.  We will show in
particular that a scale invariant picture naturally emerges (and whose
justification lies in the underlying criticality of the depinning
transition\cite{Kardar-PR98}).  It is therefore extremely important to
identify those statistical features in order to be able to up-scale
the description, and capture the transition toward a deterministic
behavior at a large enough scale.



\begin{figure}[b]
\subfigure[]{
  \centering
\includegraphics[width=.31\textwidth]{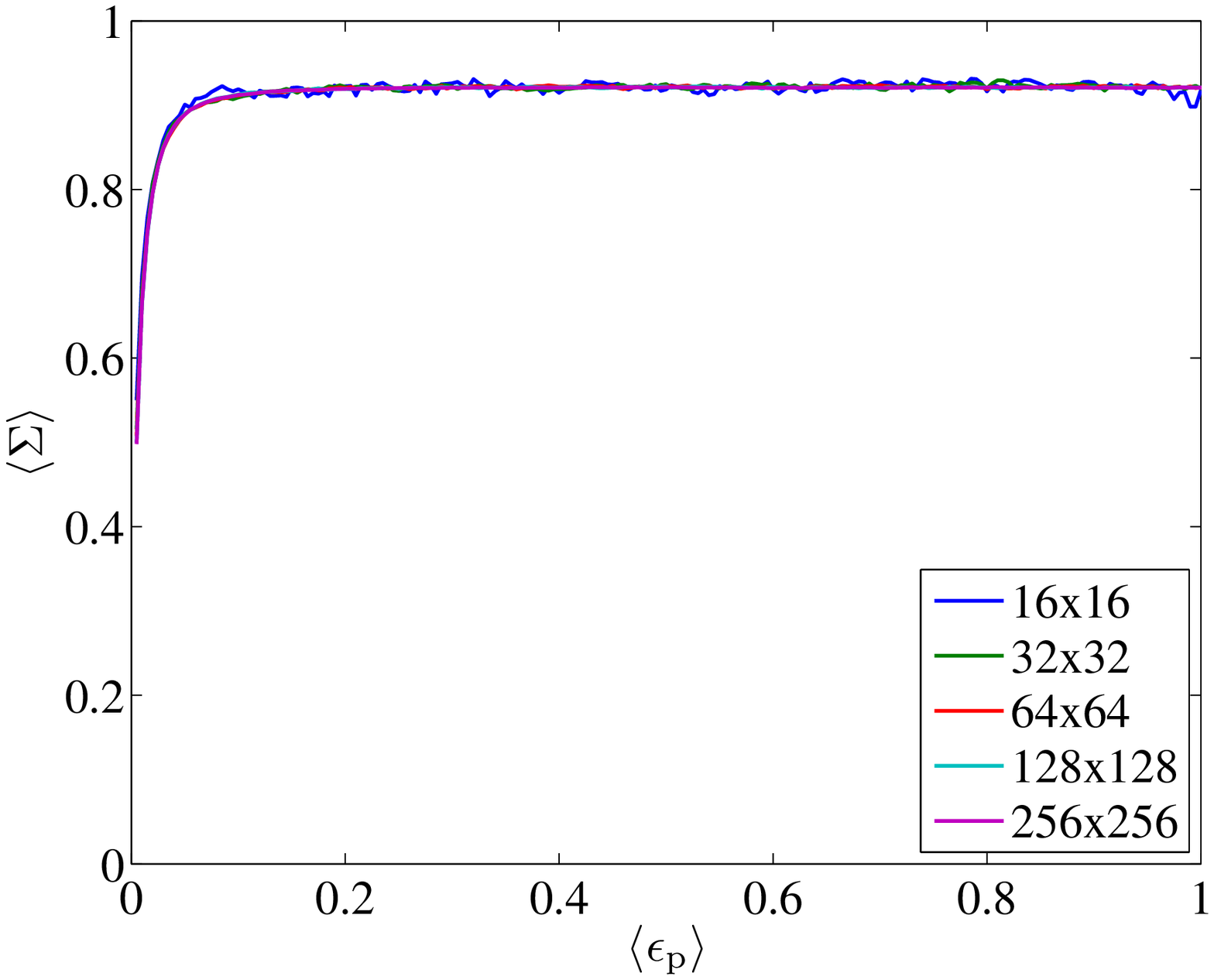}
}
\subfigure[]{
  \centering
  \includegraphics[width=.31\textwidth]{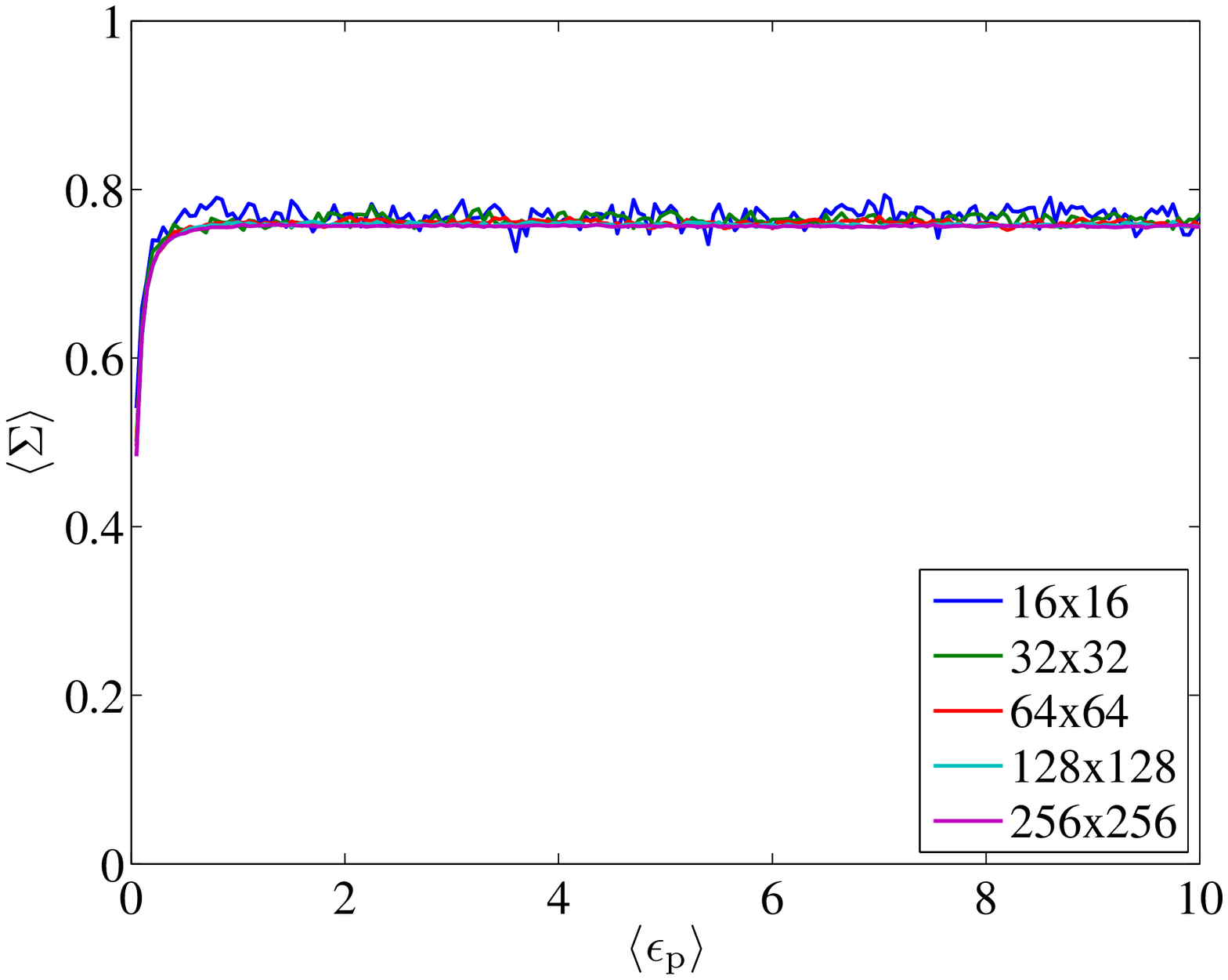}
}
\subfigure[]{
  \centering
  \includegraphics[width=.31\textwidth]{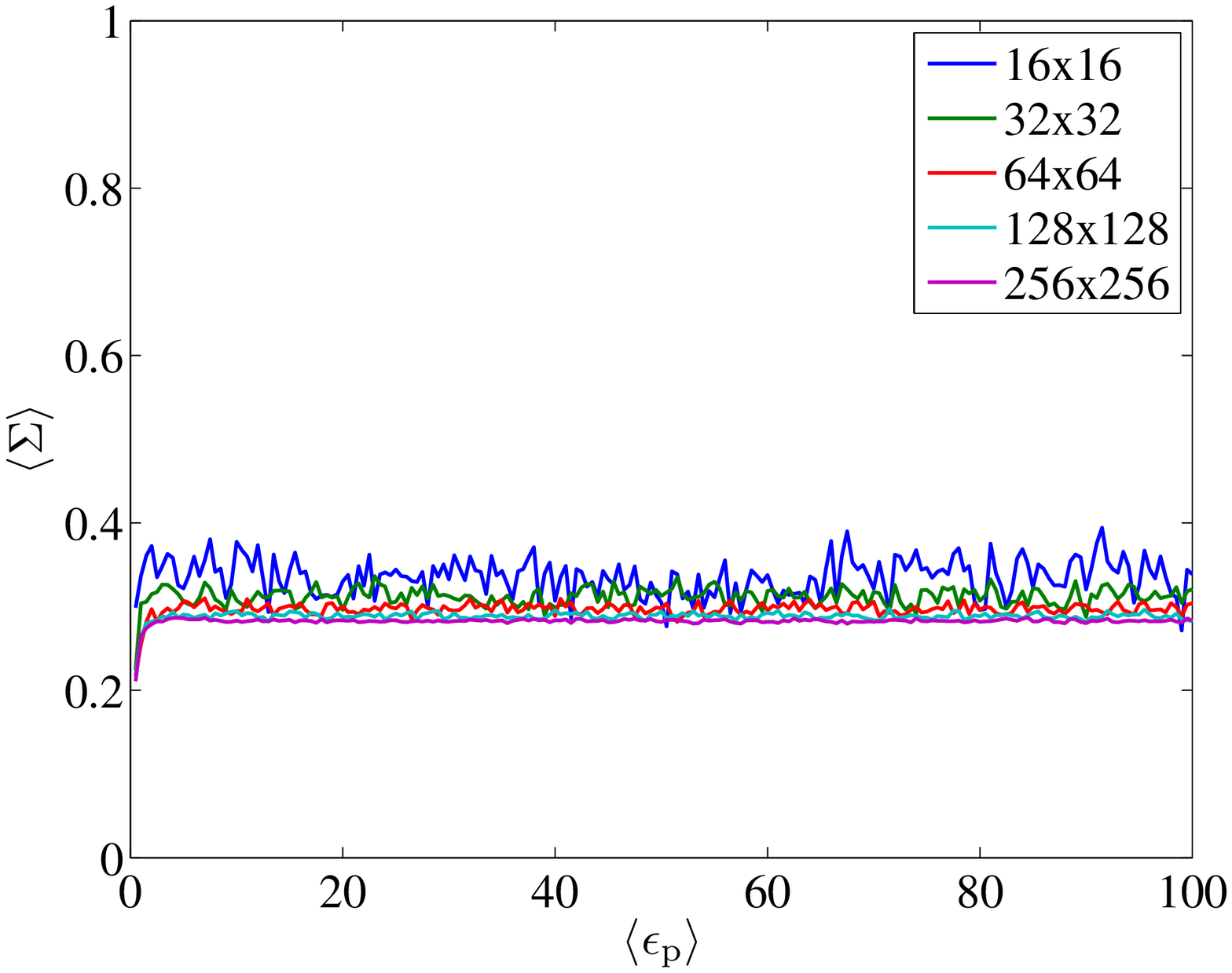}
}
  \caption{Hardening curves: Averaged stress $\langle \Sigma \rangle$
    {\it vs.} plastic strain $\varepsilon$ obtained for three values
    of the $d$ parameter.  ($d=0.01$ (a), $d=0.1$ (b) and $d=1$
    (c)). The curvature signs a transient hardening effect, {\it i.e.}
    the elastic limit (yield stress) increases with plastic
    strain. After the hardening stage, the stress saturates at a
    plateau value, the macroscopic yield stress $\Sigma_Y $. }
  \label{StsStn}
\end{figure}

\section{Stress-strain relation}\label{sec:stress-strain}

As discussed above, the choice of an extremal dynamics corresponds to
a quasistatic driving at a vanishing strain rate. While the strain
rate is kept constant, the external stress is a highly fluctuating
quantity. Far from being restrictive, this choice allows us to visit
all successive configurations of the system and to reconstruct the
response to any loading. The case of the response to a monotonically
increasing stress is recovered by computing the maximum external
stress over the past. However as this quantity is controlled by one
single event in the past, it is a fragile quantity.  In the following,
the time evolution is cut into intervals corresponding to a number of
event equal to the number of sites in the system (so that the mean
plastic strain increment is $d/2$).  The macroscopic stress is
computed as the maximum external stress over each
interval. Fig.~\ref{StsStn} shows the evolution of this quantity,
denoted here $\langle \Sigma \rangle$ as a function of the plastic
strain for different values of the slip increment parameter $d$.

It is to be noted that the stress-strain curve appears here to be a well-behaved
quantity which is independent of the system size in the explored range ($L=16$
to $L=256$ for different values of the slip increment parameter
$d=0.01\;,d=0.1\;,d=1$ in Figure~\ref{StsStn}). Saturation occurs at a
$d$-dependent value of the plateau stress.  However, as can be seen in
Figure~\ref{StsStn}, for higher values of $d$, a systematic size effect can be
seen. It affects only the smallest sizes for $d=0.1$, but appears to be very
significant for sizes ranging from 16 to 256 when $d=1$.

A simple argument allowing us to rationalize the dependence of the plateau
stress on the slip increment parameter $d$ consists of regarding the elastic
stress fluctuations as a mechanical noise comparable to a thermal one. Using
this analogy early proposed by Sollich \etal\cite{Sollich-PRL97} and more
recently discussed by Lemaitre and Caroli\cite{Lemaitre-preprint06}, each local
slip induces a stress fluctuation whose amplitude is proportional to $d$, which
can be seen accordingly as an effective mechanical temperature. Following this
interpretation, the higher $d$, the higher the mechanical noise and the lower
the external stress needed to destabilize a shear zone. A clear limit of this
interpretation is the high degree of spatial correlation of the elastic response
to a local slip. Without pursuing further this interpretation, let us simply
note that the systematic decrease of the plateau stress with the slip increment
$d$ observed in Fig.~\ref{StsStn} follows qualitatively this picture.

\begin{figure}[b]
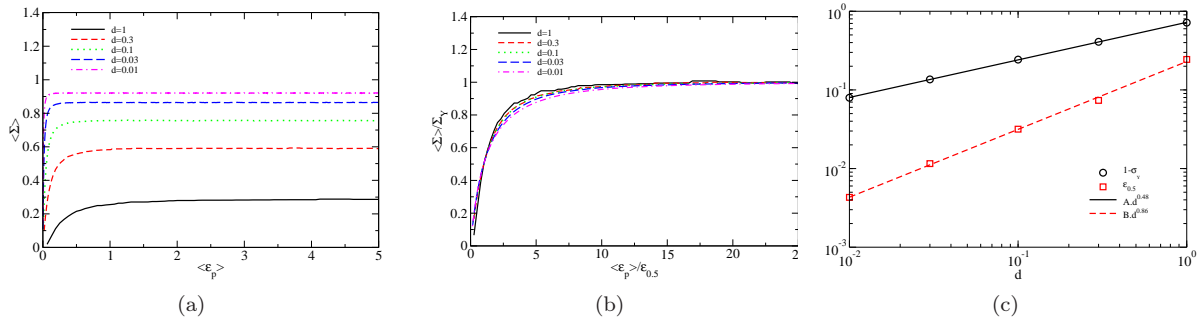

\vspace{0.5cm}
\subfigure[]{
  \centering
\includegraphics[width=.31\textwidth]{hardening256.eps}
}
\hfill
\subfigure[]{
  \centering
  \includegraphics[width=.31\textwidth]{hardening256-rescaled.eps}\hfill
}
\subfigure[]{
  \centering
  \includegraphics[width=.31\textwidth]{hardening256-scaling.eps}
}
  \caption{(a) Averaged stress $\langle\Sigma\rangle$ {\it vs.} plastic strain
    $\varepsilon$ obtained for a system of size $L=256$ with values of
    the slip increment parameter ranging from $d=0.01$ to $d=1$. The plateau stress value increases while the parameter $d$ decreases.
    (b) same data after rescaling by the plateau stress value
    $\Sigma_Y$ and the plastic strain $\varepsilon_{0.5}$ obtained at
    $\langle\Sigma\rangle=0.5\Sigma_Y$.  (c) Evolution of the plateau stress
    $\Sigma_Y$and the typical plastic strain $\varepsilon_{0.5}$ with
    the slip increment $d$.  }
  \label{StsStn-scaling}
\end{figure}

In Fig.~\ref{StsStn-scaling}a we show hardening curves obtained for a system of
size $L=256$ with values of the slip increment parameter ranging from $d=0.01$
to $d=1$. We then test in Fig. \ref{StsStn-scaling}b a simple rescaling using
the reduced variables $\langle \Sigma\rangle/\Sigma_Y$ and $\langle
\varepsilon_p \rangle/\varepsilon_{0.5} $ where $\Sigma_Y$ is the plateau stress
value and $\varepsilon_{0.5}$ is the plastic strain obtained at
$\Sigma=0.5\Sigma_Y$.  While not perfect the superimposition obtained with this
procedure is reasonable and appears to gain in quality for small values of $d$.
In Fig. \ref{StsStn-scaling}c we show the scaling behavior of the plateau stress
$\sigma_Y$ and the typical plastic strain $\varepsilon_{0.5}$. The latter shows
a sublinear dependence to $d$, $\varepsilon_{0.5}\propto d^{0.85}$ while the
former approaches unity, the upper limit of the threshold distribution with a
close to square root dependence $1-\Sigma_Y\propto d^{0.48}$.

A naive view of this behavior (leading to a square root scaling for both
quantities) consists of noting that small values of $d$ allows for a more
precise exploration of the valleys of the disordered landscape. Multiple slips
at the location are thus necessary before plastic activity jumps to another
site. Since any time, a slip occurs, the internal stress is incremented by an
elastic response of order $d$, the typical deformation can be seen as a first
return of a biased random walk of slope $d$.

The initial curvature of the stress/strain response is characteristic of a
transient hardening behavior, also called micro-plasticity {\it i.e.} the
progressive increase of the yield stress upon deformation. Such a phenomenon,
traditionally attributed to dislocation entanglement or pinning by impurities in
metal plasticity was first believed to be absent in amorphous
plasticity\cite{Schuh-ActaMat07}. Yet, the phenomenon of progressive
densification of silicate glasses upon high pressure cycling can be interpreted
as density hardening effect\cite{VDCPBCM-JPCM08}. Moreover a clear effect of
strain hardening upon shear cycling has been reported
recently in a metallic glass\cite{Schuh-APL08}. Within the present model, it is possible to give a
simple statistical interpretation of strain hardening phenomenon in absence of
dislocations. As early discussed in Ref.  \cite{BVR-PRL02} the increase of the
yield stress can be related to the progressive exhaustion of the weakest spots
of the materials.

\begin{figure}[t]
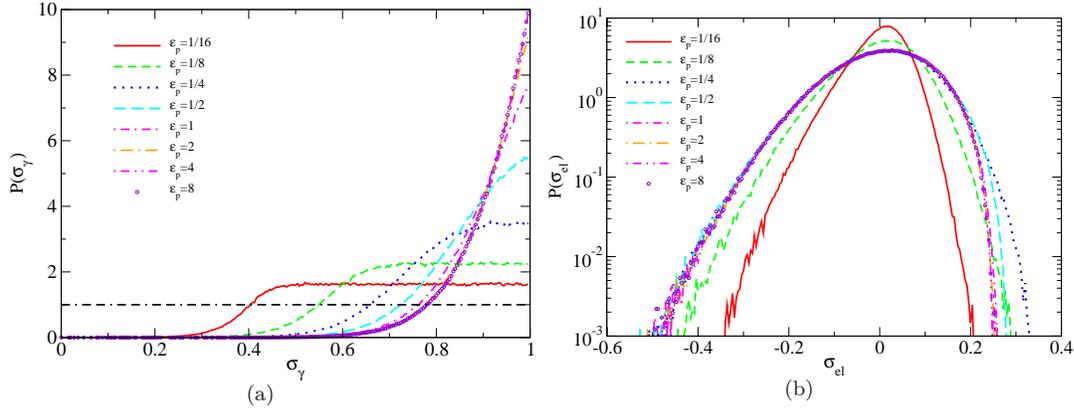

\subfigure[]{
  \label{fig:mini:subfig:a}
  \begin{minipage}[h]{0.425\textwidth}
   \centering
   \includegraphics[width=70mm]{histo_trap.eps}
  \end{minipage}
  \label{ThdDU}}
 \subfigure[]{
  \label{fig:mini:subfig:b}
  \begin{minipage}[h]{0.425\textwidth}
   \centering
   \includegraphics[width=70mm]{histo_stress.eps}
  \end{minipage}
  \label{ThdDGD}}
  \caption{Evolution of (a) the distribution of microscopic plastic
    thresholds and (b) the distribution of internal stress under
    increasing plastic strain (from $\varepsilon_p=1/16$ to
    $\varepsilon_p=8$). Data are obtained with a system of size $L=64$
    with a slip increment parameter $d=0.1$, averaged over 250
    realizations. The horizontal black dot-dashed line gives the
    distribution of thresholds in the initial state
    ($\varepsilon_p=0$). Due to the exhaustion of weak sites during
    deformation, the distribution of thresholds progressively shifts
    toward high values when $\varepsilon_p$ inncreases and finally
    converges toward a limit distribution of typical width $d$. The
    internal stress distribution (here represented in semi-log
    coordinates) converges toward a Gaussian-like distribution with
    however a notable asymmetry: a fatter tail for the negative
    stress.}
  \label{ThdD}
\end{figure}

This phenomenon is clearly visible in Fig. \ref{ThdD} where we
represented the evolution under shear of the distribution of local
thresholds $P(\sigma_\gamma)$ and of the distribution of residual
elastic stress $P(\sigma_{el})$. While originally uniform in the
interval $[0,1]$, the distribution of plastic thresholds is
progressively shifted toward high values by exhaustion of the weakest
sites before reaching a limit steady state distribution. In the
transient regime, a self-organized-critical-like dynamics seems at
work: weak sites are replaced by normal ones, thus inducing a
systematic bias to the distribution. While less remarkable a similar
trend may be extracted from the evolution of the stress
distribution. A clear asymmetry is visible: large positive stress
values which favor local reorganization are less frequent than their
negative counterparts.

\section{A non-persistent localizing behavior}\label{sec:localization}

By the very definition of the model, the mean plastic strain increases in
proportion to the number of events, or ``time''.  As the mean plastic strain
increases, the macroscopic stress reaches a plateau in the steady state,
$\langle \rmd \sigma\rangle/\rmd \varepsilon=0$. Using classical results from
continuum mechanics, for homogeneous media, the system is at the limit between a
hardening behavior $\rmd \sigma/\rmd \varepsilon>0$, expected to give rise to a
homogeneous plastic strain, and a softening regime $\rmd \sigma/\rmd
\varepsilon<0$ where strain should localize.  A simple way to illustrate this
property, is to note that a uniform (but of arbitrary magnitude) slip along a
line oriented at $\pm \pi/4$ with respect to the principal axes, will not induce
any stress in the medium.  Introducing $\bm n_1=1/\sqrt{2}(1,1)$ and $\bm
n_2=1/\sqrt{2}(1,-1)$, and arbitrary scalar functions $f$ and $g$,
    \be\label{eq:silent_modes}
    \varepsilon_p(\bm x)=f(\bm x.\bm n_1)+g(\bm x.\bm n_2)
    \ee
is a field which does not involve any elastic stress within the medium. In our
model, the residual elastic stress is the only way to endow the system with
memory, and hence, no limiting (nor amplifying) mechanism will act against
(with) such degrees of freedom.

\subsection{Fluctuations of plastic strain and stress}

\begin{figure}[b]
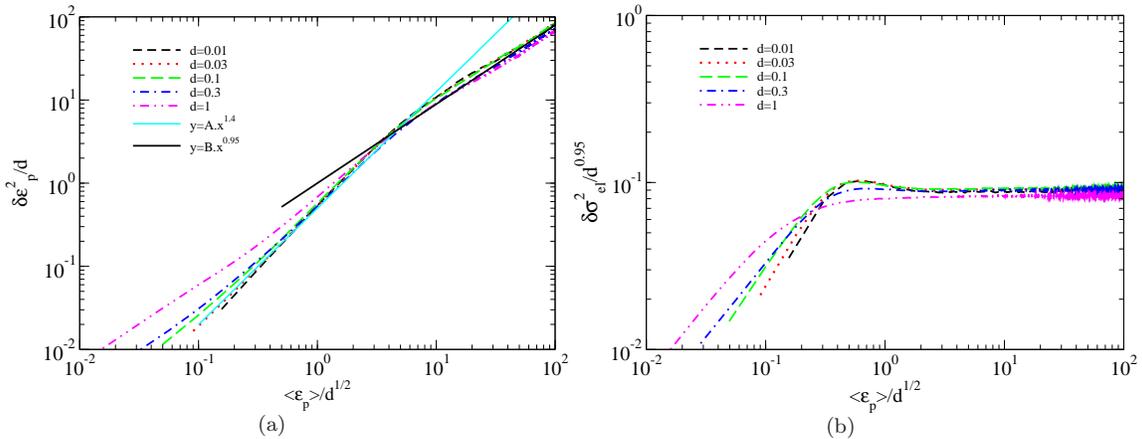

\vspace{0.5cm}
  \centering
\subfigure[]{
  \label{fig:mini:subfig:a}
  \begin{minipage}[h]{0.45\textwidth}
   \centering
   \includegraphics[width=75mm]{strain-width-rescaled.eps}
  \end{minipage}
  \label{width-strain}}
 \subfigure[]{
  \label{fig:mini:subfig:b}
  \begin{minipage}[h]{0.45\textwidth}
   \centering
   \includegraphics[width=75mm]{stress-width-rescaled.eps}
  \end{minipage}
 \label{width-stress}}
  \caption{(a) Log-log plot of the evolution of the rescaled variance
    of plastic strain $\delta \varepsilon_p^2/d$ with rescaled plastic
    strain $\langle\varepsilon_p\rangle/d^{0.5}$ ($L=128$,
    $d=0.01,\;0.03,\;0.1,\;0.3,\;1$). A blue (or light-gray) straight
    line of slope $1.4$ is shown as a reference in the transient
    regime, and a black straight line of slope 0.95 is shown in the
    ``stationary'' regime.  (b) Evolution of the rescaled residual
    stress standard deviation $\delta \sigma_{el}/d^{0.95}$ in the
    same coordinates. In contrast with strain, a real saturation
    regime can be identified.}
  \label{roughness}
\end{figure}

The presence of disorder gives rise to stress fluctuations around the average,
which may affect this simple picture, and hence it is of interest to
characterize the inhomogeneity of the plastic strain field.  A first simple
estimator is the variance of the entire plastic strain field,
\be
\delta\varepsilon_p^2=[\langle\varepsilon_p^2\rangle-\langle\varepsilon_p\rangle^2]
\ee
which is to be characterized as a function of the mean plastic strain
$\langle\varepsilon_p\rangle$. Indeed, on the one hand, for a plastic
strain field which does not display any long range correlation,
$\delta\varepsilon_p^2$, should saturate to a constant, independent of
$\langle\varepsilon_p\rangle$ and $L$. If, on the other hand, the
plastic strain is localized on a simple slip system (a straight line
spanning the entire system), then along this localization band, namely
for $L$ sites, the plastic strain amounts to
$\varepsilon_1=L\langle\varepsilon_p\rangle$, whereas it is null over
the rest of the system.  In this case, the variance amounts to
$\delta\varepsilon_p^2\propto L\langle\varepsilon_p\rangle^2$. Therefore
the scaling of $\delta\varepsilon_p$ with $L$ and $\langle
\varepsilon_p\rangle$ is informative on the more or less uniform
distribution of plastic strain.

The evolution of the variance of the plastic strain as a function of the mean
strain is shown in Fig.~\ref{roughness}a in a log-log plot for systems of size
$L=128$ and various values of the slip increment parameter ranging from $d=0.01$
to $d=1$. A reasonable collapse onto a master curve can be obtained when using
the reduced plastic strain $ \varepsilon_p/d^{1/2}$. As discussed above the
quality of the rescaling procedure increases when the value of $d$ decreases.

A transient power-law like regime can be identified. An indicative
straight line of slope 1.4 is shown in the transient regime. Then the
strain fluctuation seems to slowly transit toward a diffusive like
regime of slope close to unity. Again an indicative straight line of
slope 0.95 is shown in this regime.
Coming back to the above discussion we thus obtain a power law
evolution of the variance of strain $\delta\varepsilon_p^2 \propto
\langle \varepsilon_p\rangle^\alpha$ with an exponent $\alpha\approx
1.4$ in the transient regime and $\alpha\approx 1$ in a second
diffusive-like regime.  These results are thus intermediate between a
homogeneous deformation regime ($\alpha=0)$ and a full localization
behavior ($\alpha=2)$. 

In Fig.~\ref{roughness}b we represent the
evolution of the elastic stress variance. The contrast with plastic
strain fluctuation is striking since we observe here a clear
saturation after the transient regime. This means that plastic strain
is dominated by stress free soft modes $\pm 45^\circ$. These results
help us to give a first picture of the behavior of the system. After
spatial correlations have been established in a transient localized
regime (high value of the exponent $\alpha$), the system transits
toward a diffusive regime where plastic deformation can be seen as a
random succession of shear bands in the directions at $\pm 45^\circ$.

\subsection{Maps of plastic strain field}

\begin{figure}[htbp]
  \centering
   \includegraphics[width=50mm]{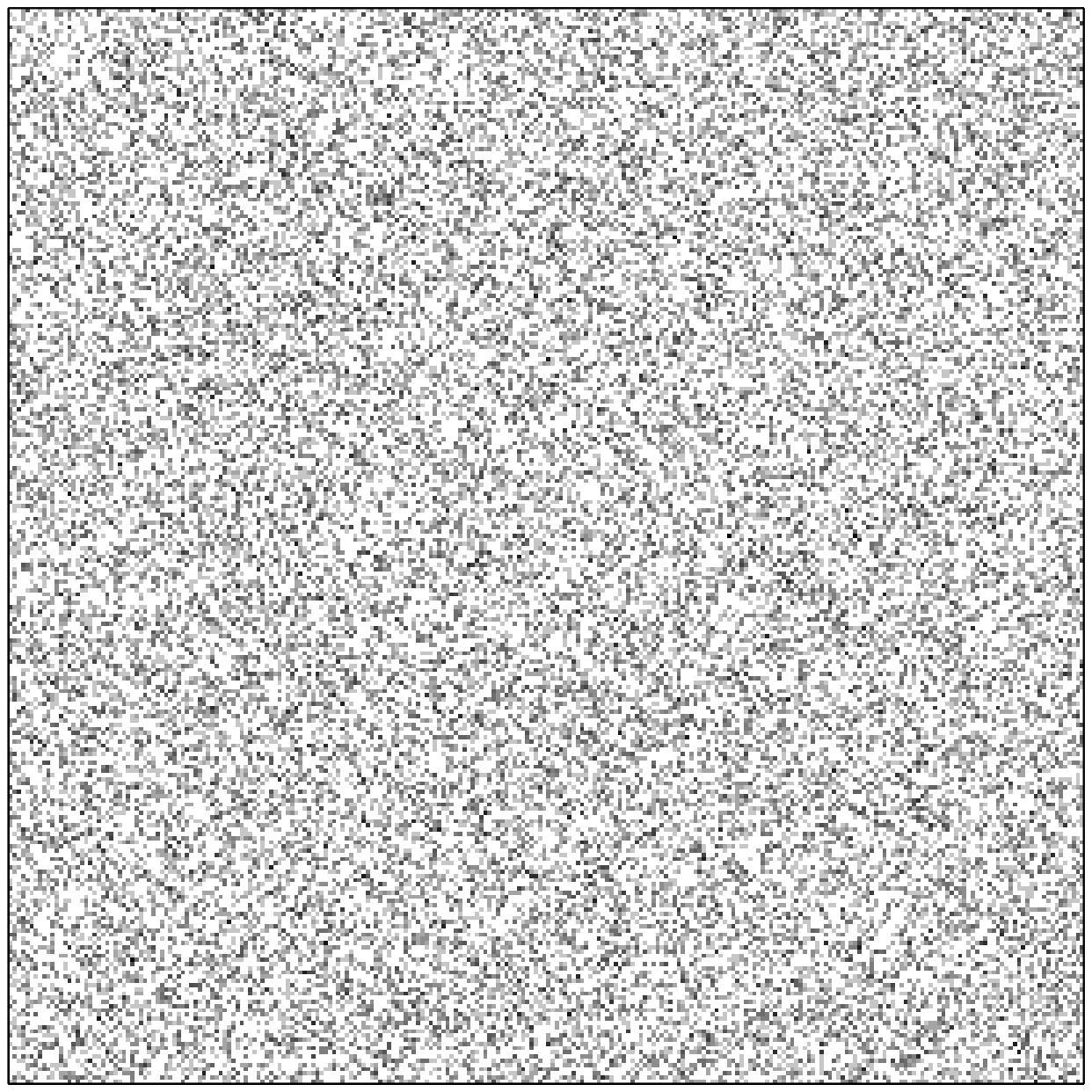}
   \includegraphics[width=50mm]{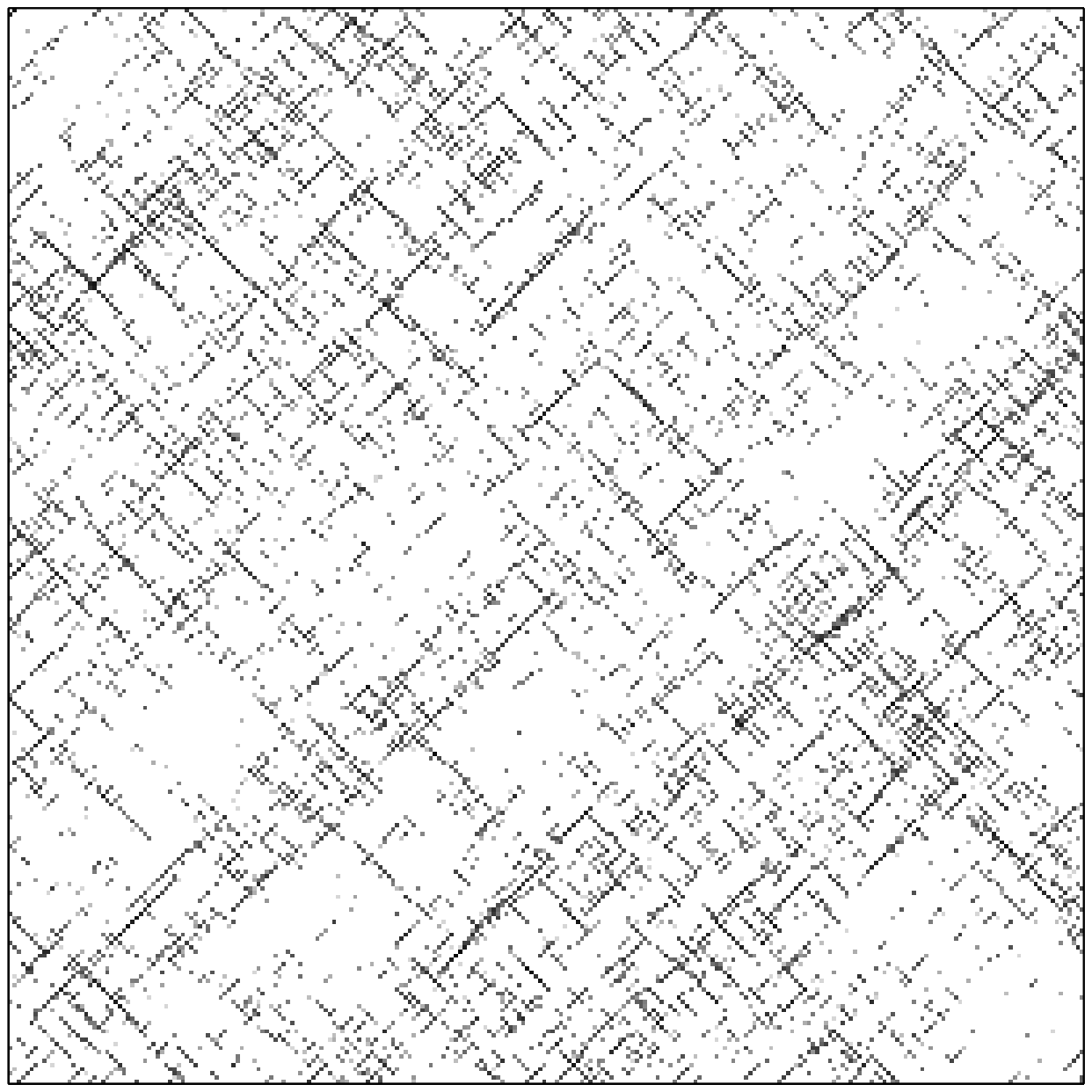}
   \includegraphics[width=50mm]{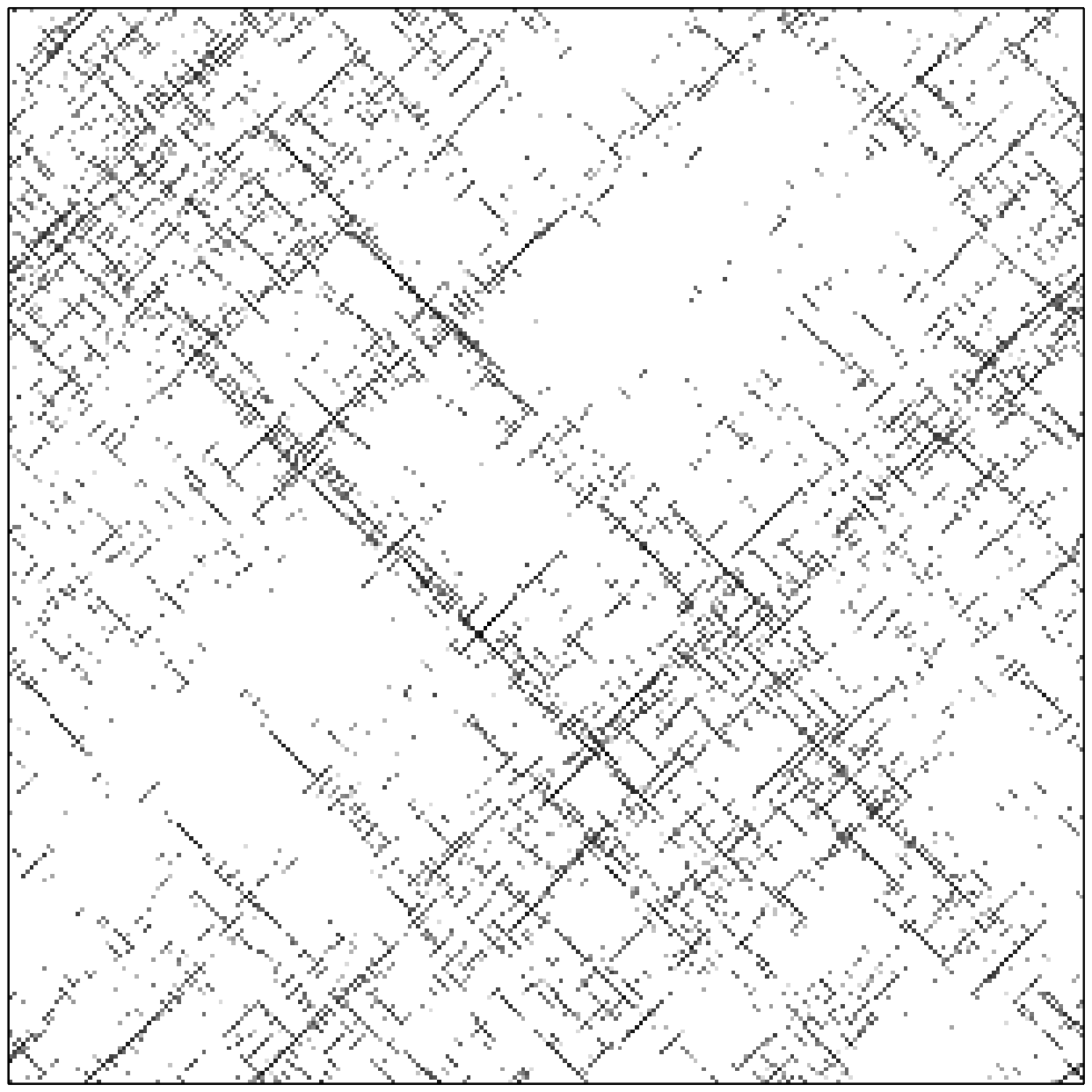}\\
   \includegraphics[width=50mm]{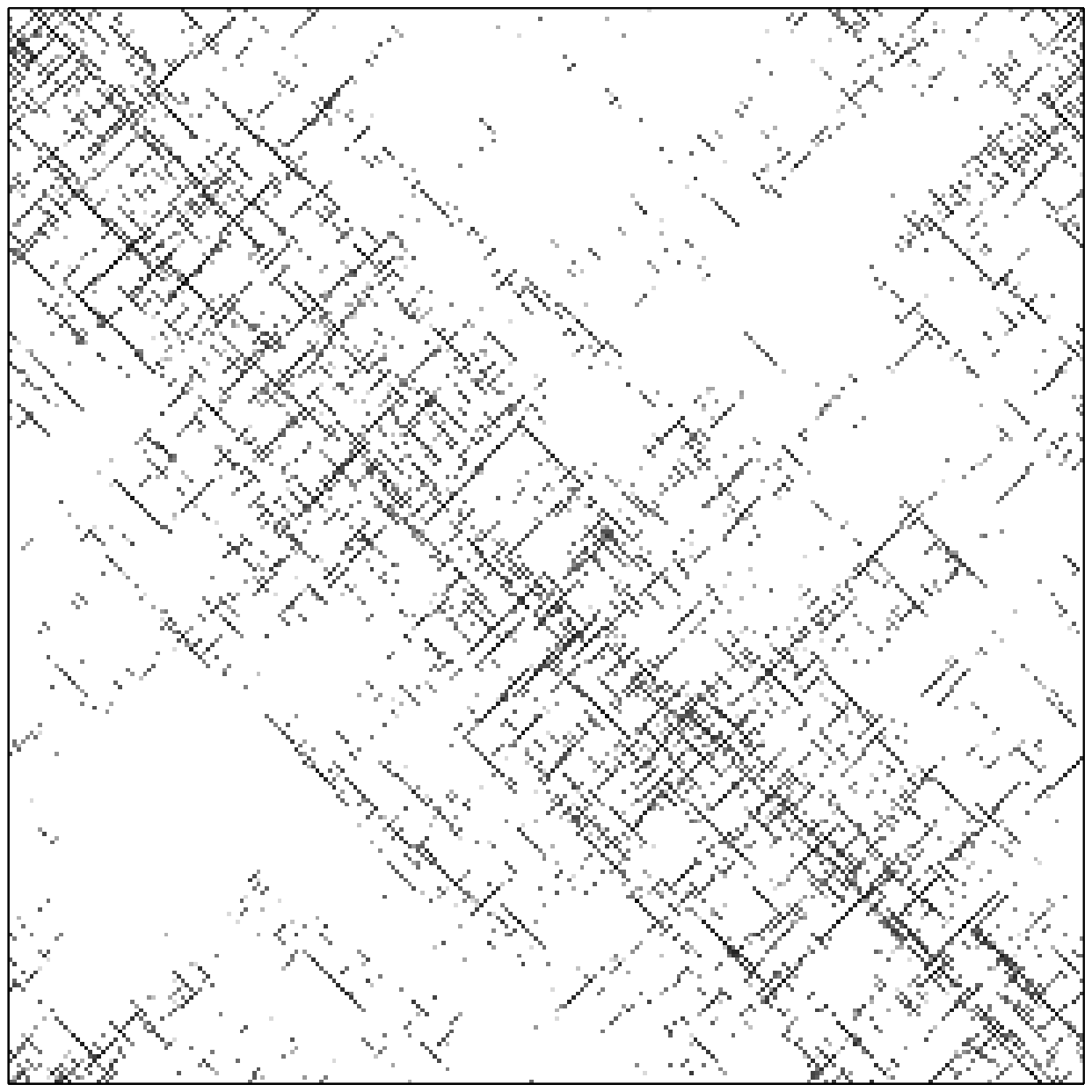}
   \includegraphics[width=50mm]{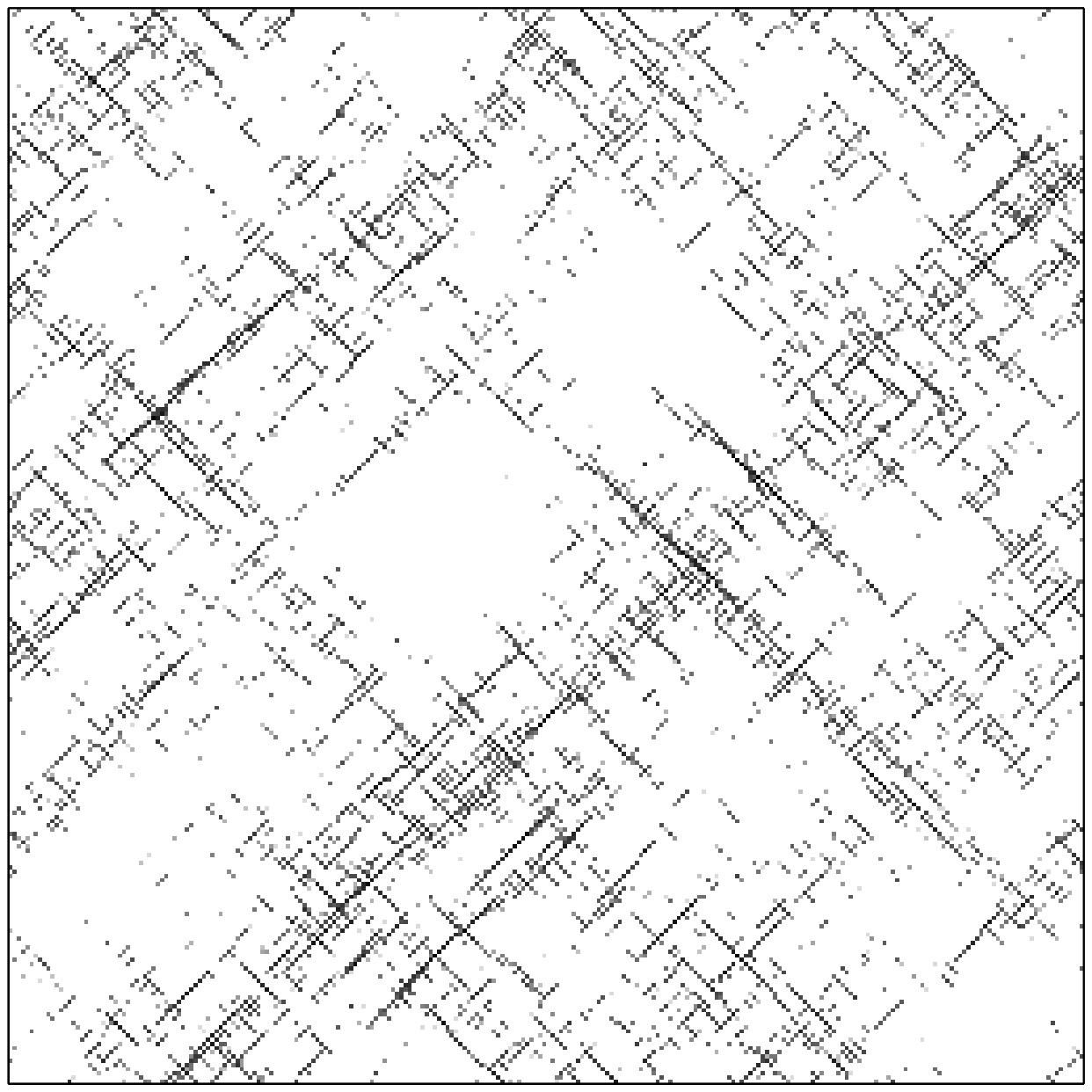}
   \includegraphics[width=50mm]{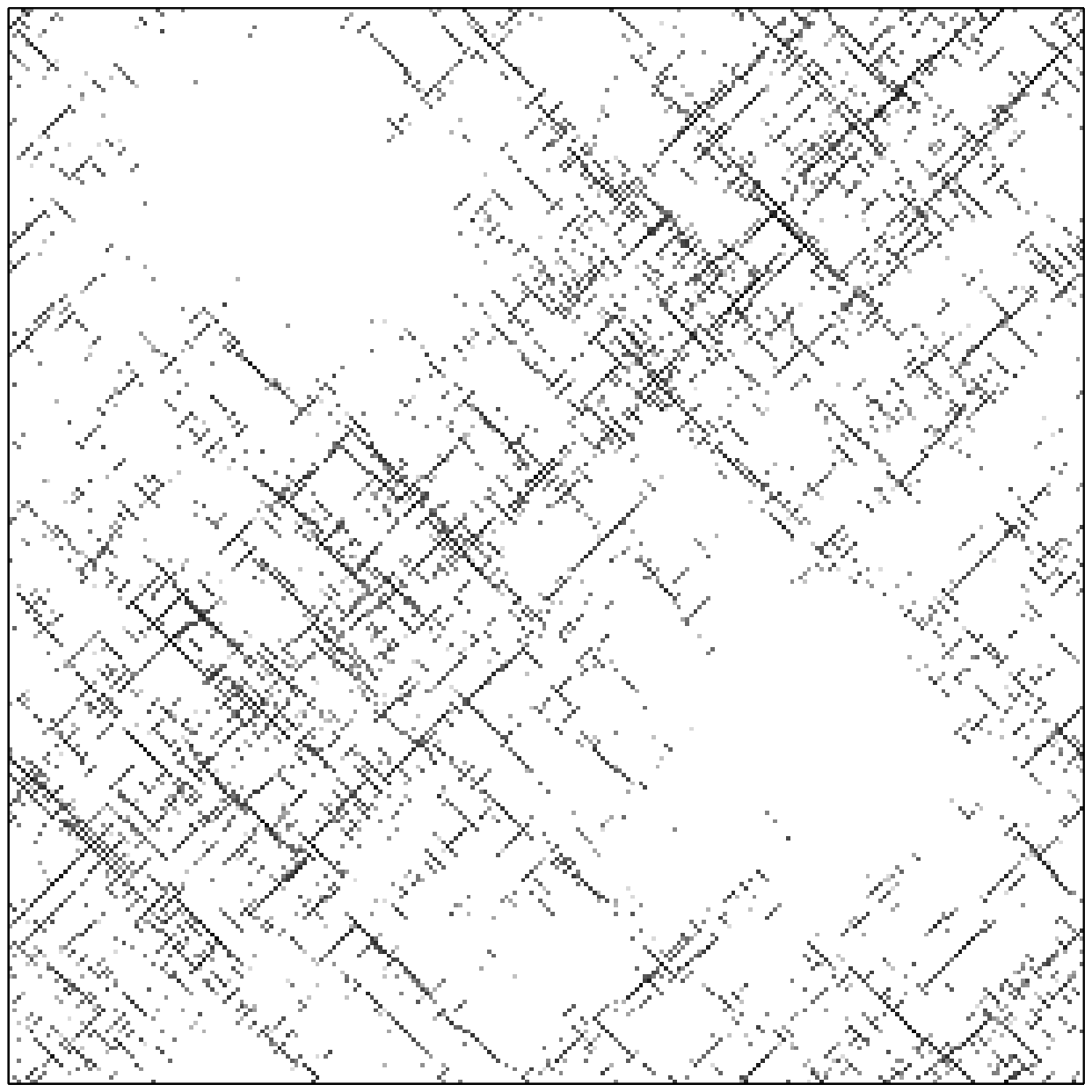}
\\
\vspace{2mm}\hspace{-57mm}
   \includegraphics[width=50mm]{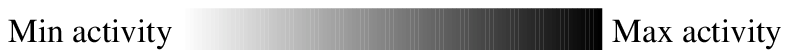}
  \caption{Maps of cumulated plastic activity  during
    a finite strain window  $\Delta \varepsilon_p = 0.01$ corresponding to increasing levels of
    deformation: $\varepsilon_p=0.01,0.10,0.20,0.30,0.50,1.00$. The
    system size is $L=256$. While random at the beginning, the plastic
    activity progressively localizes along $\bm n_1$ and $\bm n_2$ directions.}
  \label{EvtPM}
\end{figure}

Beyond this macroscopic behavior, a closer look at the local scale
gives evidence for the development of strong spatial
correlations. Fig. \ref{EvtPM} presents a series of maps of plastic
activity at different levels of mean plastic strain. Every map
corresponds to the plastic strain accumulated during a finite strain
window $\Delta \varepsilon_p = 0.01$. While the plastic activity is
initially homogeneously distributed in space, a progressive
development of correlations along the $\bm n_1$ and $\bm n_2$
directions can be observed. Simultaneously, it appears that larger and
larger regions remain still, i.e. without any plastic flow.  As
earlier noted, plastic strain appears as streaks aligned with either
$\bm n_1$ or $\bm n_2$.  Let us note that these patterns observed at
the largest scale generate little stresses as they can grossly be
described by the modes described in Eq.~\ref{eq:silent_modes}.

\begin{figure}[htbp]
\subfigure[]{
  \centering
  \includegraphics[width=50mm]{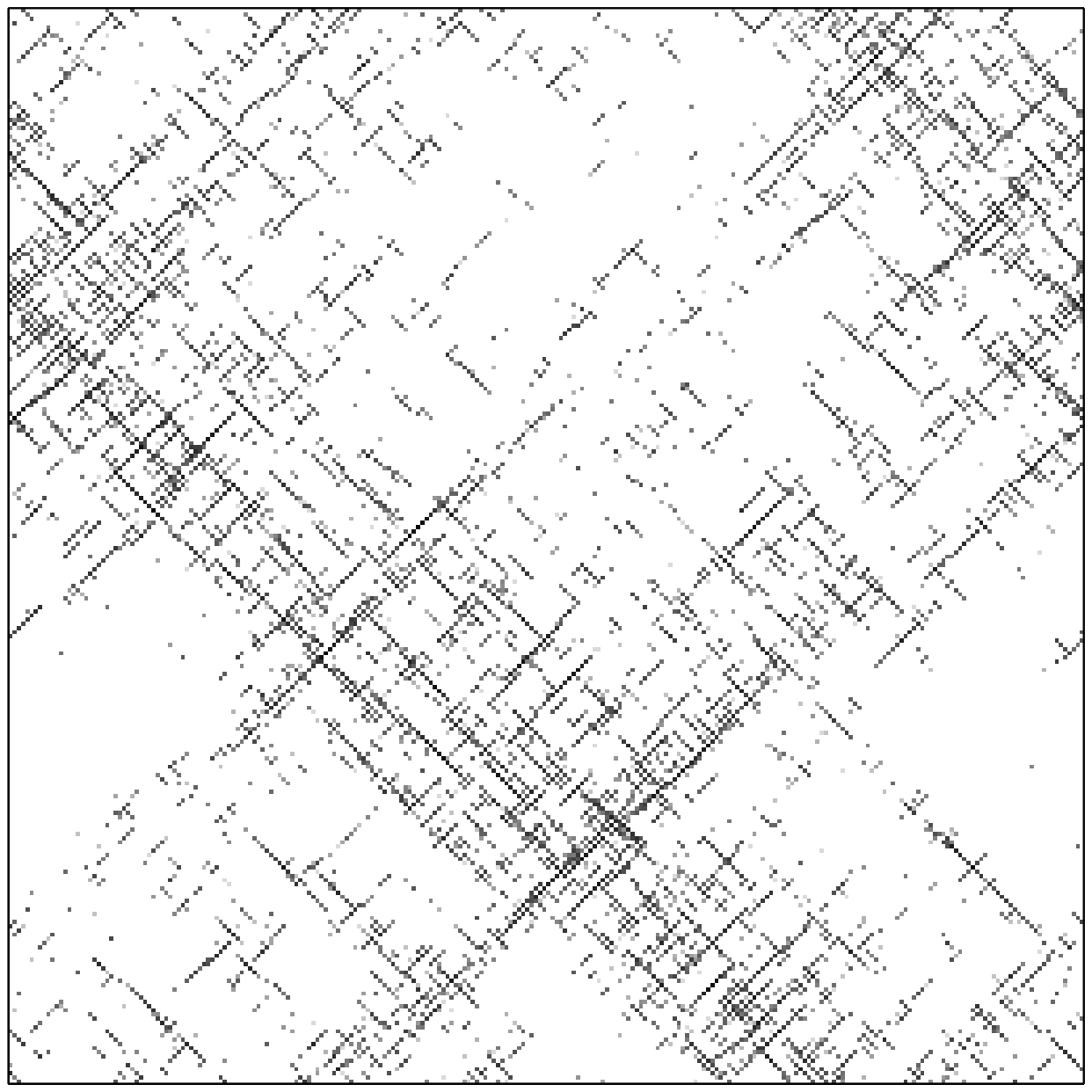}
}
\subfigure[]{
  \centering
  \includegraphics[width=50mm]{EvtP256100.eps}
}
\subfigure[]{
  \centering
  \includegraphics[width=49.2mm]{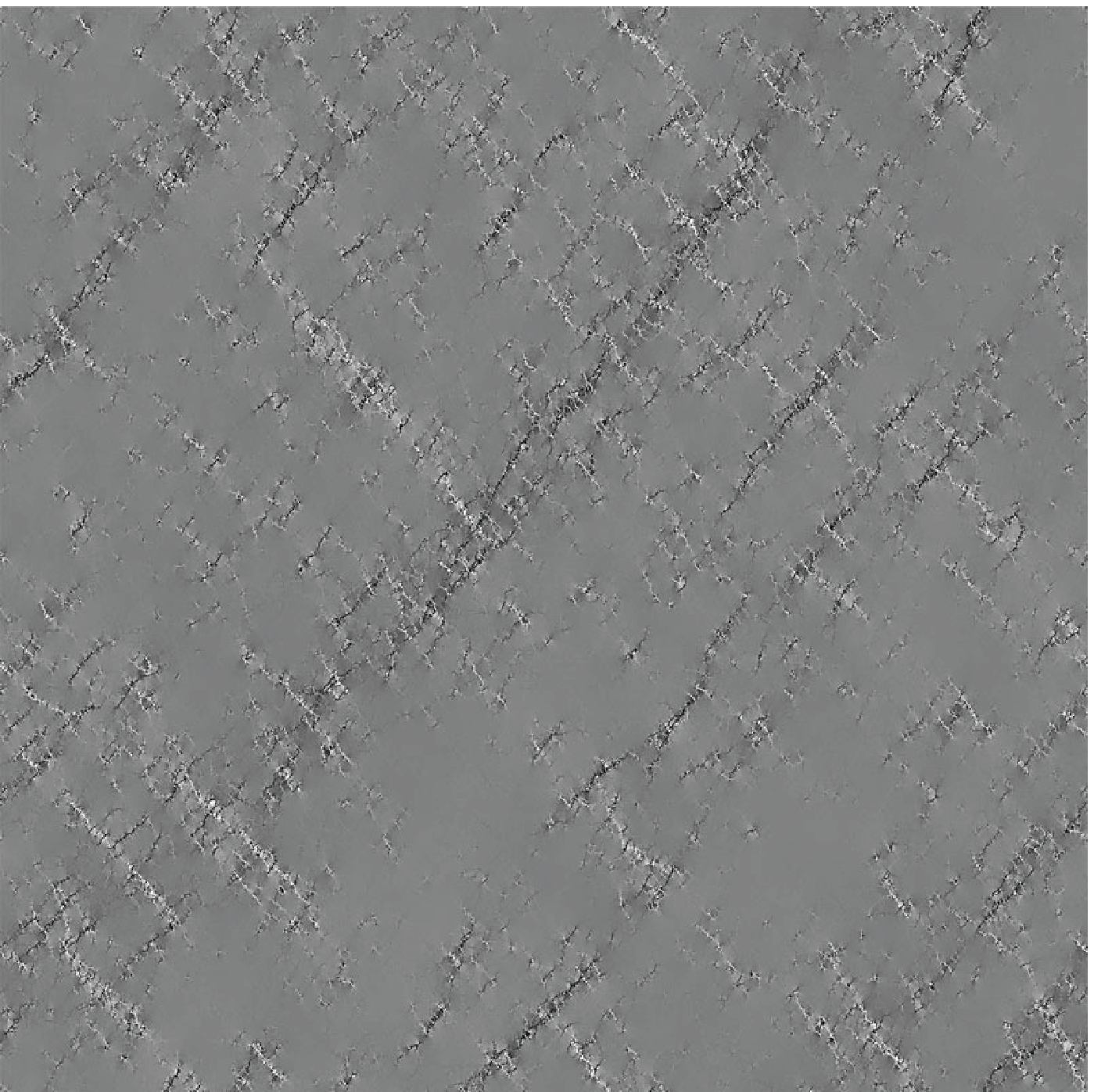}
}
\\
  \caption{(a) Map of cumulated plastic activity during a finite
    strain window $\Delta \varepsilon_p = 0.01$ taken at
    $\varepsilon=0.95$ with $L=256$, $d=0.01$.  A clear localization
    of the plastic deformation is observed. (b) same as (a) but
    taken ``later'' at $\varepsilon_p=1.0$. Again plasticity localizes
    along elongated structures at $\pm 45^\circ$ but the patterns are
    markedly different: localization is not persistent. (c) for
    comparison, reproduction of a strikingly similar map of plastic
    activity (vorticity of the displacement field) recently obtained
    by Maloney and Robbins\cite{Maloney-PRL09}) on a 2D Lennard-Jones
    glass under compression.}
  \label{EvtPNP}
\end{figure}

Note that this localization appears not to be persistent. While the incremental
plastic strain displays similar patterns at different times, shear bands are not
superimposed, but they rather move through the system with statistically similar
features past the initial transient at different deformation levels; they seem
to diffuse in the system. The two left panels of Fig. \ref{EvtPNP} where we
represented the plastic activity at $\varepsilon_p=0.95$ and $\varepsilon_p=1.0$
give an illustration of this competition between localization and diffusion. In
the second map, we recover hints of the first one but most of the plastic
activity has moved elsewhere. The last panel of Fig. \ref{EvtPNP} proposes a
striking comparison with recent numerical results by Maloney and
Robbins\cite{Maloney-JPCM08,Maloney-PRL09}. These authors study the behavior of
a two-dimensional Lennard-Jones glass under uni-axial compression, and periodic
boundary condition for the stress and strain field, a situation which is quite
comparable to our own boundary conditions. They show the vorticity of the
displacement field, in order to highlight the zones of concentrated plastic
strain.  With this representation, a slip line along the $\bm n_1$ (resp. $\bm
n_2$) direction appears as a line of concentrated negative (resp. positive)
vorticity, whereas the local density of plastic strain does not distinguish
between both directions.

The growth of the correlation length takes place in the initial transient up to
the stage where it reaches the system size, and steady state is established.  Up
to now, we essentially focused on this initial stage.  However, since we
observed the formation of large scale structures, it is of interest to
characterize the signature of the largest structures in the steady state.  In
the following sections space correlations in the stress and plastic strain
fields are discussed.




\section{Anisotropic plastic strain correlations}\label{sec:strain_correlation}

\begin{figure}[htbp]
\centering
\subfigure[]{
  \begin{minipage}[h]{0.45\textwidth}
   \centering
\includegraphics[width=60mm]{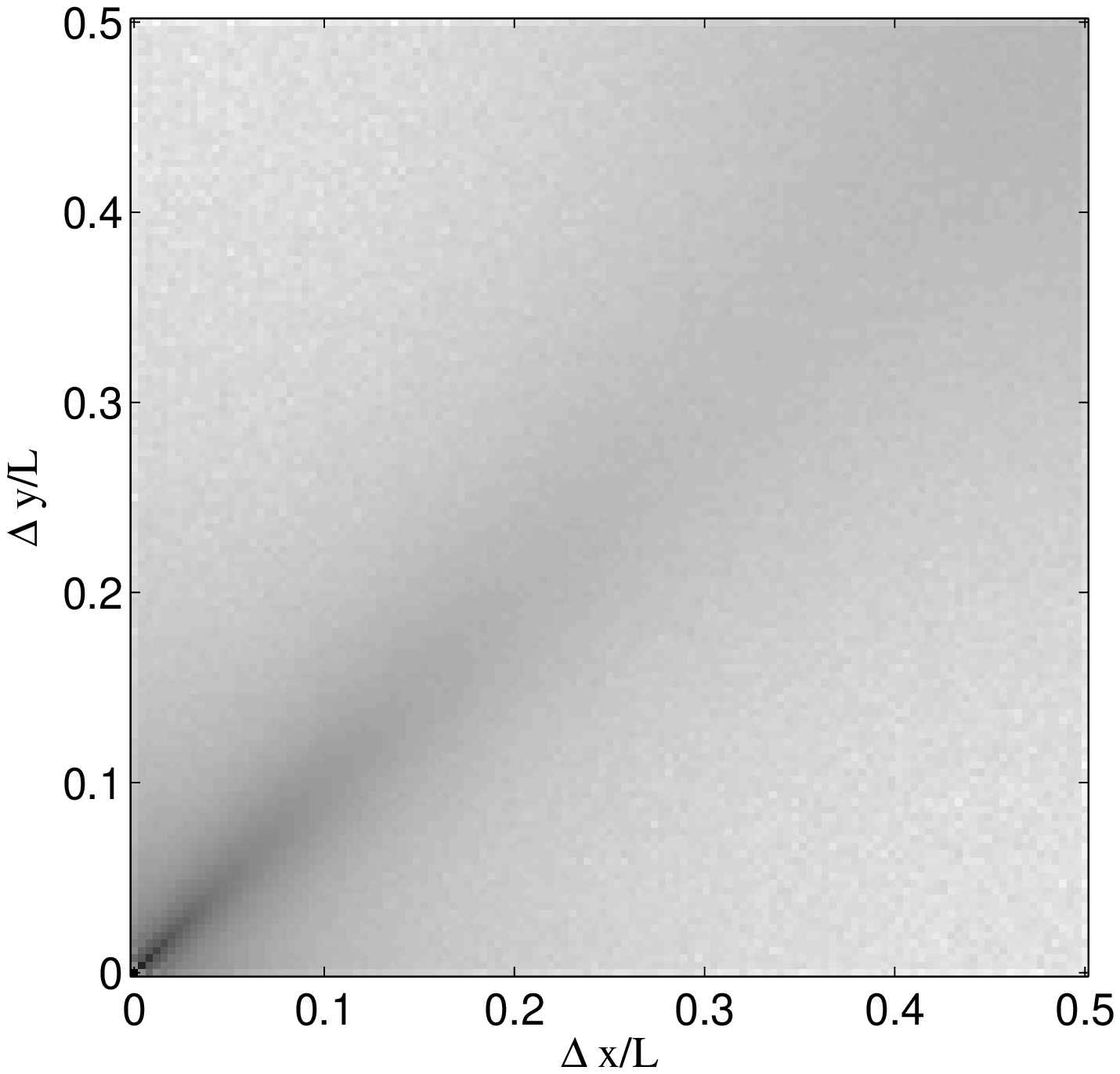}\\
\hspace{-18mm}
 \includegraphics[width=50mm]{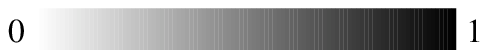}
\vspace{2mm}
  \end{minipage}
  \label{Pdydx_a}}
 \subfigure[]{
  \begin{minipage}[h]{0.42\textwidth}
   \centering
   \includegraphics[width=80mm]{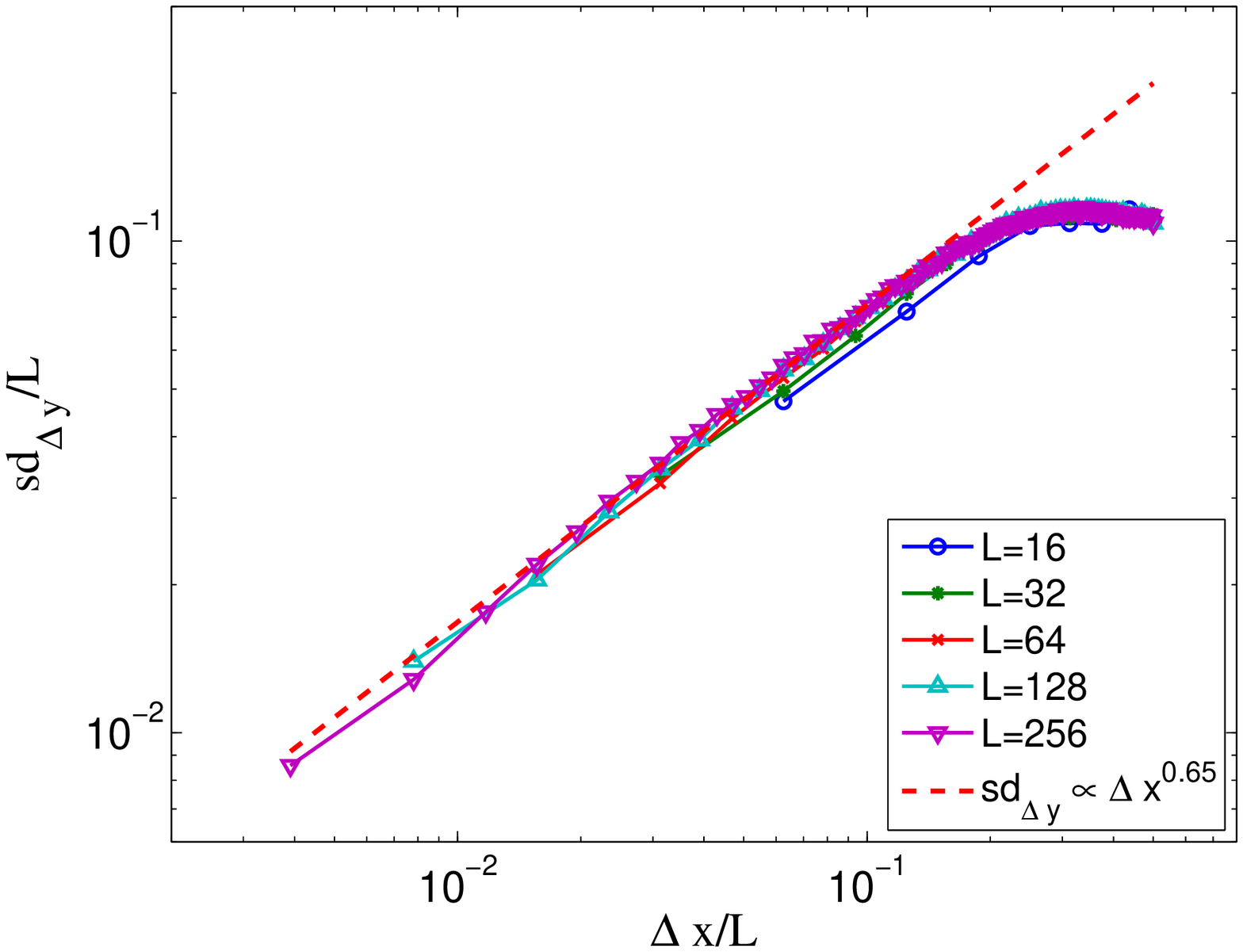}
  \end{minipage}
  \label{Pdydx_b}}
  \caption{(a) Map of $P(\Delta x,\Delta y)$, the probability that two
    successive plastic events are separated by distances $\Delta x$ and
    $\Delta y$ along $x-$ and $y$-axis for a system of size $L=256$.
    Preferential angles at $\pm\pi/4$ are clearly visible. (b) Standard
    deviation $sd_{\Delta y}$ of the separation of two consecutive
    plastic events along the $y-$axis for a given value of the
    separation distance $\Delta x$. A power law behavior $sd_{\Delta
    y} \propto \Delta x^{\zeta}$ is obtained with $\zeta \approx
    0.65$. The fact that $\zeta<1$ shows that the relative width of
    those shear bands tends to 0 for large system sizes. }
  \label{Pdydx}
\end{figure}

We now give a quantitative analysis of the anisotropic plastic strain
correlations. A first method consists in considering the spatial distribution of
successive plastic events $P(\Delta x, \Delta y)$. A map of this distribution is
presented in Fig.~\ref{Pdydx_a} for the upper right quadrant.  Without surprise,
we obtain a panache that obeys the quadrupolar symmetry of the elastic kernel: a
plastic event is far more probable in a direction at $\pm \pi/4$ degree than
along the principal axis. The width of this panache can be quantified.
Fig.~\ref{Pdydx_b} shows the standard deviation $sd_{\Delta y}$ of the distance
$\Delta y$ for a given value of the distance $\Delta x$ between two successive
plastic events. We obtain a nice scaling relation $sd_{\Delta y}\propto \Delta
x^\zeta$ where $\zeta\approx 0.65$. Note that a similar value was obtained in
the anti-plane shear case discussed in Ref.~\cite{BVR-PRL02}.  Such a value
corresponds also to the roughness exponent of minimal path in a random
potential~\cite{Kardar-PRL87} which can easily be proven to be relevant for
anti-plane plasticity with quenched random yield stress. The fact that the value
of this exponent is less than unity indicates an important property of the
plastic strain patterns: the aspect ratio decreases as $\Delta x^{\zeta-1} $ so
that asymptotically, shear is concentrated on straight bands of vanishing
relative width.

\begin{figure}[t]
\subfigure[]{
  \centering
\includegraphics[width=.35\textwidth]{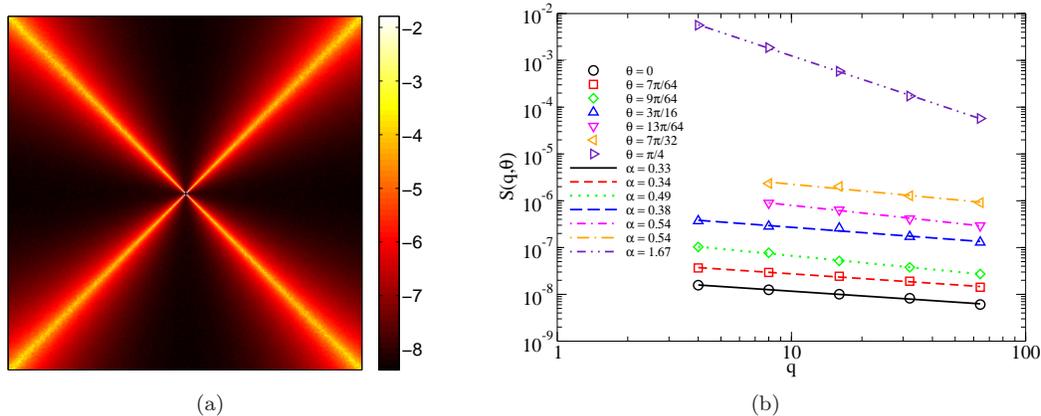}
}
\hspace{0.5cm}
\subfigure[]{
  \centering
\includegraphics[width=.45\textwidth]{Pspec02bis.eps}
}
  \caption{(a) Power spectrum of the plastic strain using a
    decimal log scale. A clear
    quadrupolar symmetry is obtained. (b) Cuts along the axis at
    $\theta=0$ and $\theta=\pi/4$. Different power law behaviors
    $S(q)\propto q^{-\alpha}$ are obtained between the extreme cases
    $\alpha_0\approx 0.33$ and $\alpha_{\pi/4}\approx 1.67$.
    $L=256$ and $d=0.01$.}
  \label{PowSpec}
\end{figure}

In order to characterize the correlations of the plastic strain field,
the most natural analysis is to compute the power spectrum of the
plastic strain field.  In Fig.~\ref{PowSpec}a, an ensemble
average of 2D power spectra of the strain field is shown. This map,
here encoded in logarithmic scale, exhibits a marked fourfold
symmetry, with obvious preferential directions along $\bm n_1$ and
$\bm n_2$. To be more quantitative, we now consider the scaling of
power spectra cuts along the preferred directions and along the
principal axes, as shown in Fig.~\ref{PowSpec}b.  Both of these
spectra can be characterized by a power-law behavior with exponents
$\alpha_{\pi/4}\approx 1.7$, $\alpha_0\approx0.3$ where the indices
denote the orientation of the wave vector with respect to the
$x$-axis. Such behaviors can equivalently be characterized as
self-affine, with respective roughness exponents
$\zeta_{\pi/4}\approx0.35$ and $\zeta_{0}\approx-0.35$.  It is however
to be emphasized that the most important feature of power spectra
along directions $\theta\neq\pm \pi/4$, is that their magnitude is
orders of magnitude lower than for $\theta\pm \pi/4$.

In the case of anti-plane shear, a previous study~\cite{BVR-PRL02} also revealed
an anisotropic plastic strain field in the steady state. In that case, the power
spectra of $\varepsilon_p$ showed a different scaling with the wave-number
parallel or perpendicular to the orientation of the shear band.

In their recent study of a Lennard-Jones glass under
compression~\cite{Maloney-JPCM08,Maloney-PRL09}, Maloney and Robbins discussed
the anisotropic scaling of plastic strain. As shown in Fig. \ref{EvtPNP}, when
representing the vorticity of the displacement field, they obtained maps of
plastic activity strikingly similar to ours. Looking at the correlation of the
vorticity they obtained a power spectrum with a four-fold symmetry:
  \be
    S(q,\theta) \propto
    a(\theta) q^{-\alpha(\theta)} \;,\qquad \alpha(\theta)= 0.68
    -0.5\cos(4\theta)
    \label{scaling-MB}
  \ee
and thus in particular $\alpha_{\pi/4}=1.18$ and $\alpha_0=0.18$, to compare to
the above reported values.  In the latter expression, $a(\theta)$ is
unspecified.

\begin{figure}[htbp]
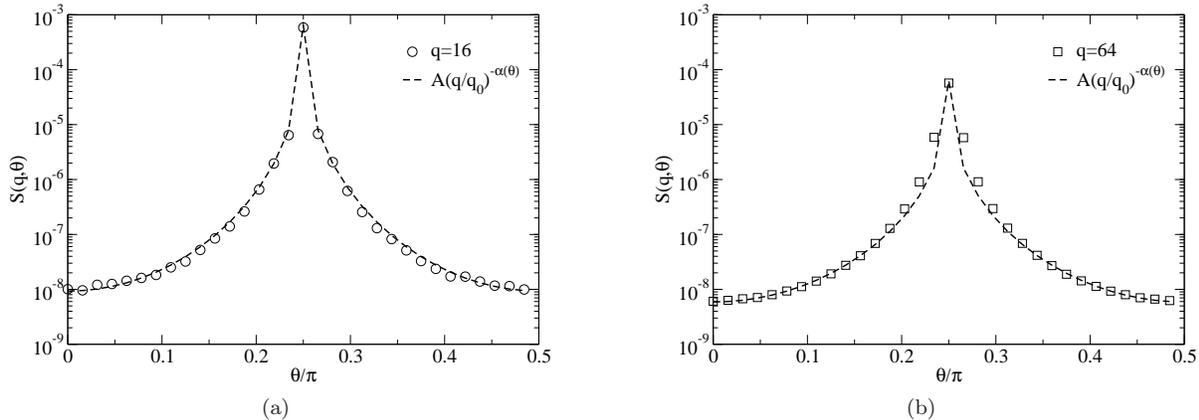


\subfigure[]{
  \centering
\includegraphics[width=.45\textwidth]{S16.eps}
}
\hfill
\subfigure[]{
  \centering
\includegraphics[width=.45\textwidth]{S64.eps}
}
  \caption{Angular dependence of the power spectrum obtained for two
    values of the wavenumber: $q=16$ (a) and $q=64$ (b) with
    the parameters $L=256$, $d=0.01$. The symbols represent the
    numerical data; The dashed lines corresponds to the best fit
    obtained with a scaling form $s(q,\theta) \propto A
    \left(q/q_0\right)^{-\alpha(\theta)}$ with an exponent
    $\alpha(\theta)= 1.67-1.34 |\cos(2\theta)|^{0.4}$ exhibiting a
    quadrupolar symmetry.}
  \label{Spec-Ang}
\end{figure}

Considering the power spectra of the cumulated plastic strain, we expect to
observe a directly comparable quantity. Indeed, we do observe a direction
dependent scaling exponent. In Fig. \ref{PowSpec} we reported the scaling
behavior obtained along cuts in different directions while the angular
dependence of $S(q,\theta)$ is shown in Fig. \ref{Spec-Ang}. Again most of the
plastic activity appears to be concentrated along the directions at $\pm \pi/4$.
Except in the close vicinity of that preferred direction, the value of the
scaling exponent remains close to $\alpha_0$, and only shows a slow increase
when the direction of the cut approaches the diagonal.

Representing part of the same results as a function of the polar angle, we
observe that an equation comparable to the previous, gives a good account of our
data
  \be
    S(q,\theta) = A
    \left(q/q_0\right)^{-\alpha(\theta)}\;, \quad \alpha(\theta)=
    \alpha_{\pi/4}-(\alpha_{\pi/4}-\alpha_0) |\cos(2\theta)|^{0.4}
    \label{scaling-TPRV1}
  \ee
where we considered the case $d=0.01$. The angular dependency of the power
spectrum is written in a slightly different manner, but both share the same
fourfold symmetry.  At a fixed wavevector modulus, Eq.~\ref{scaling-TPRV1}
proposes an amplitude of $A q_0^{\alpha(\theta)}$ whereas Maloney and Robbins
propose $A(\theta)$. However the difference is marginal, as the lower exponents
correspond to much smaller amplitudes.

\section{Conclusions}\label{sec:conclusion}

The plasticity of amorphous media was addressed through a meso-scale
modeling in order to highlight the large scale mechanical
behavior. This modeling was constructed in order to retain the most
important features of a macroscopic description and compared favorably
with much richer descriptions.  The macroscopic behavior tends to
approach an elastic/perfectly plastic law.  The progressive approach
to a constant yield stress is interpreted as a statistical selection
of high threshold local configuration.  In the perfect plasticity
regime, strain tends to localize along the direction of maximum shear
as could have been anticipated. However, the fact that elementary
plastic events are localized in space give rise to specific features
in the spatial correlation of plastic strains (with self-similar power
spectra), and fluctuations of macroscopic yield stress occurring at
large scales. Let us finally note that this model has recently been
shown to exhibit a non-trivial avalanche behavior\cite{TPVR-PRE11};
moreover the influence of the initial state of the medium has been
shown to favor an early localization of the strain onto a unique and
persistent shear band\cite{VR-PRB11}.

An extension accounting for the coupling between deviatoric and volumetric
strain remains to be investigated. It would also allow to study the hardening
behavior while paying attention to the potential apparition of an anisotropic
texture as recently discussed in \cite{RVTBR-PRL09}.


\bigskip
{\sl  We are thankful to C.E. Maloney and M.O. Robbins for the authorization
of replicating their figure in Fig.~\ref{EvtPNP} from
Ref.~\cite{Maloney-PRL09}.}

\bibliographystyle{elsarticle-num}

\end{document}

\end{document}

\section{Influence of initial state}

While the richness of the physics of depinning models mainly relies on the
competition between elasticity and disorder, we see here that the anisotropic
character and the abundance of soft modes in the elastic interaction which
characterize the present model of amorphous plasticity, naturally induce an
additional competition between localization and disorder.

\begin{figure}[htbp]
  \centering
   \includegraphics[width=50mm]{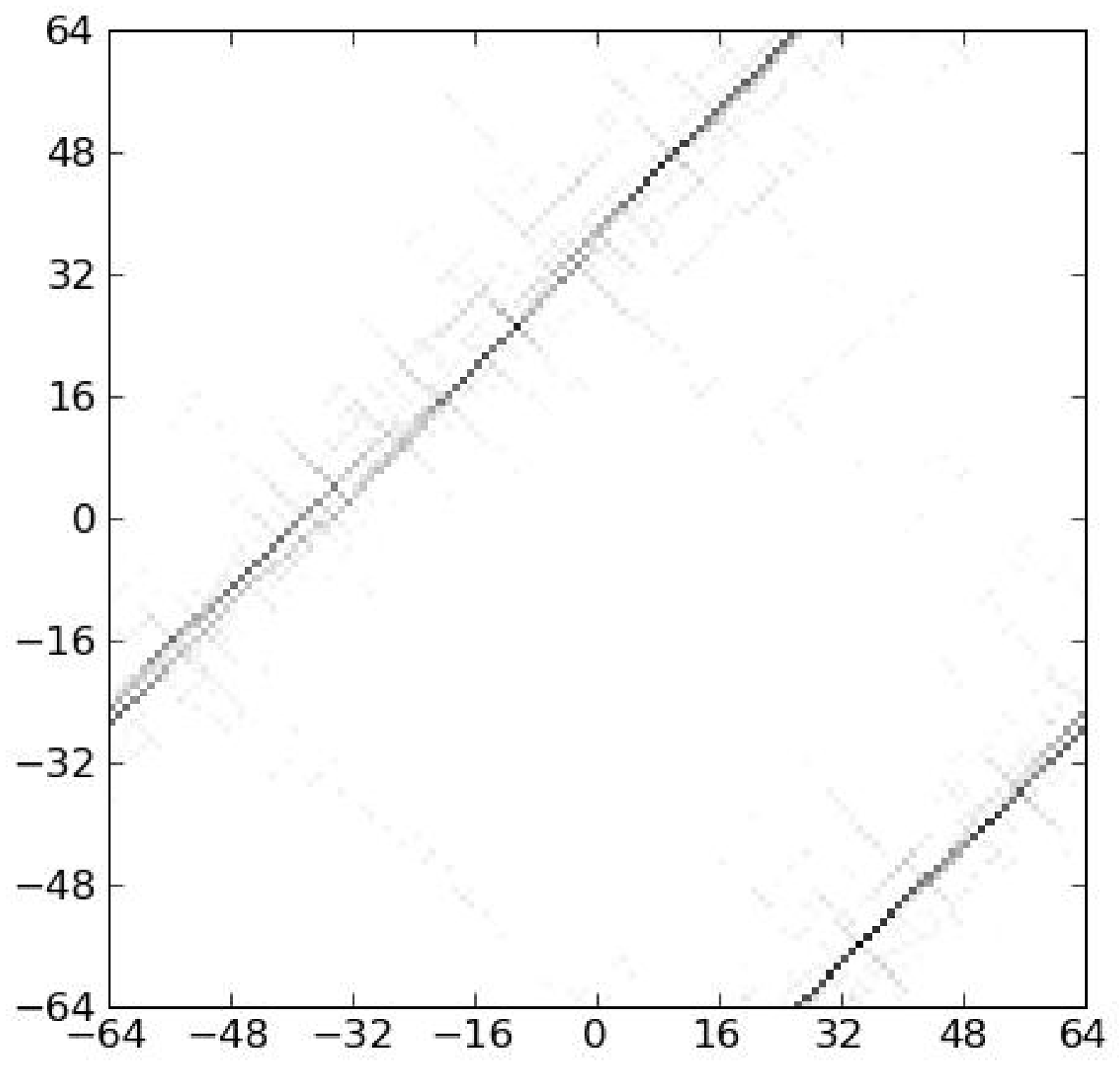}
   \includegraphics[width=50mm]{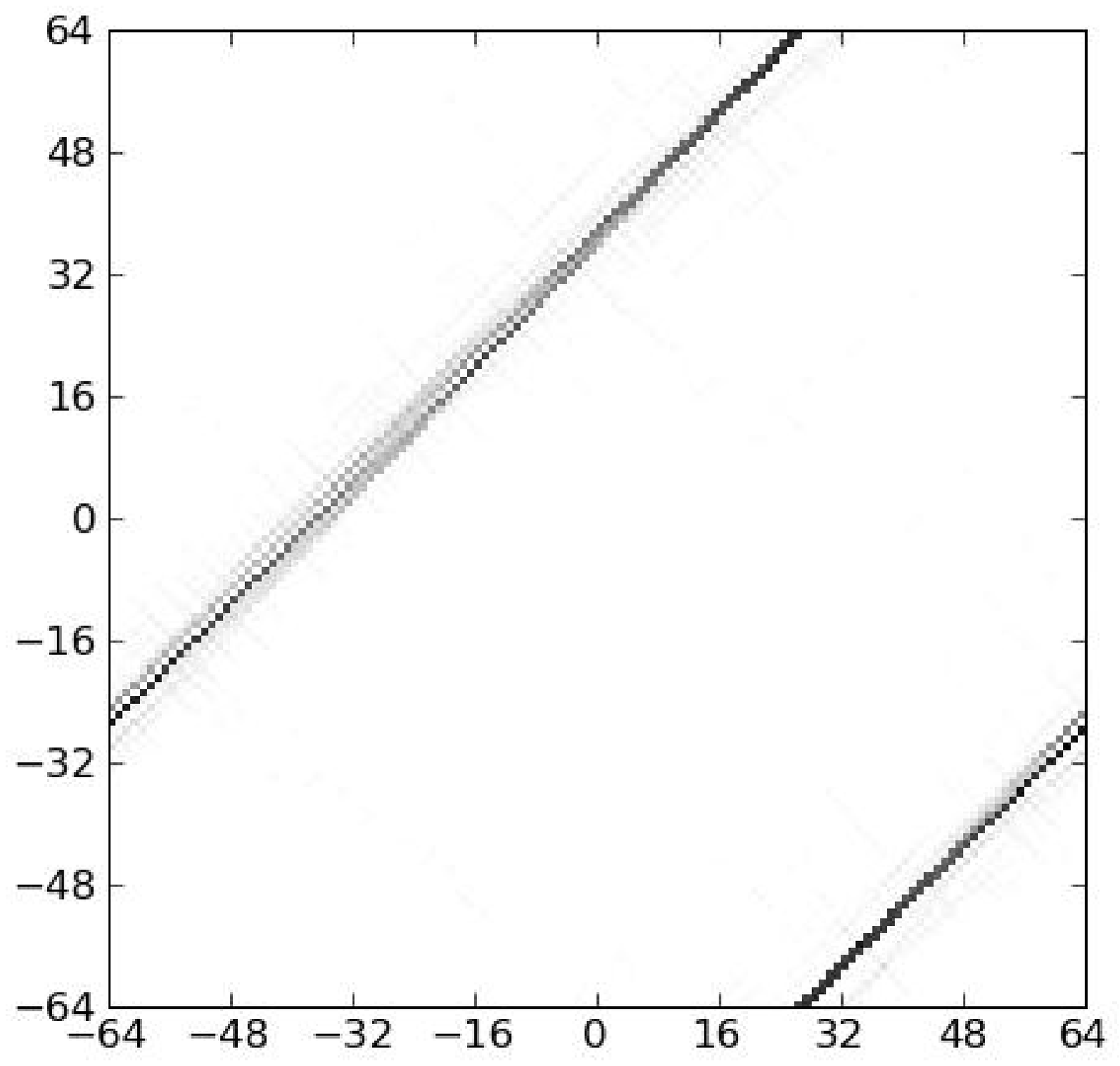}
   \includegraphics[width=50mm]{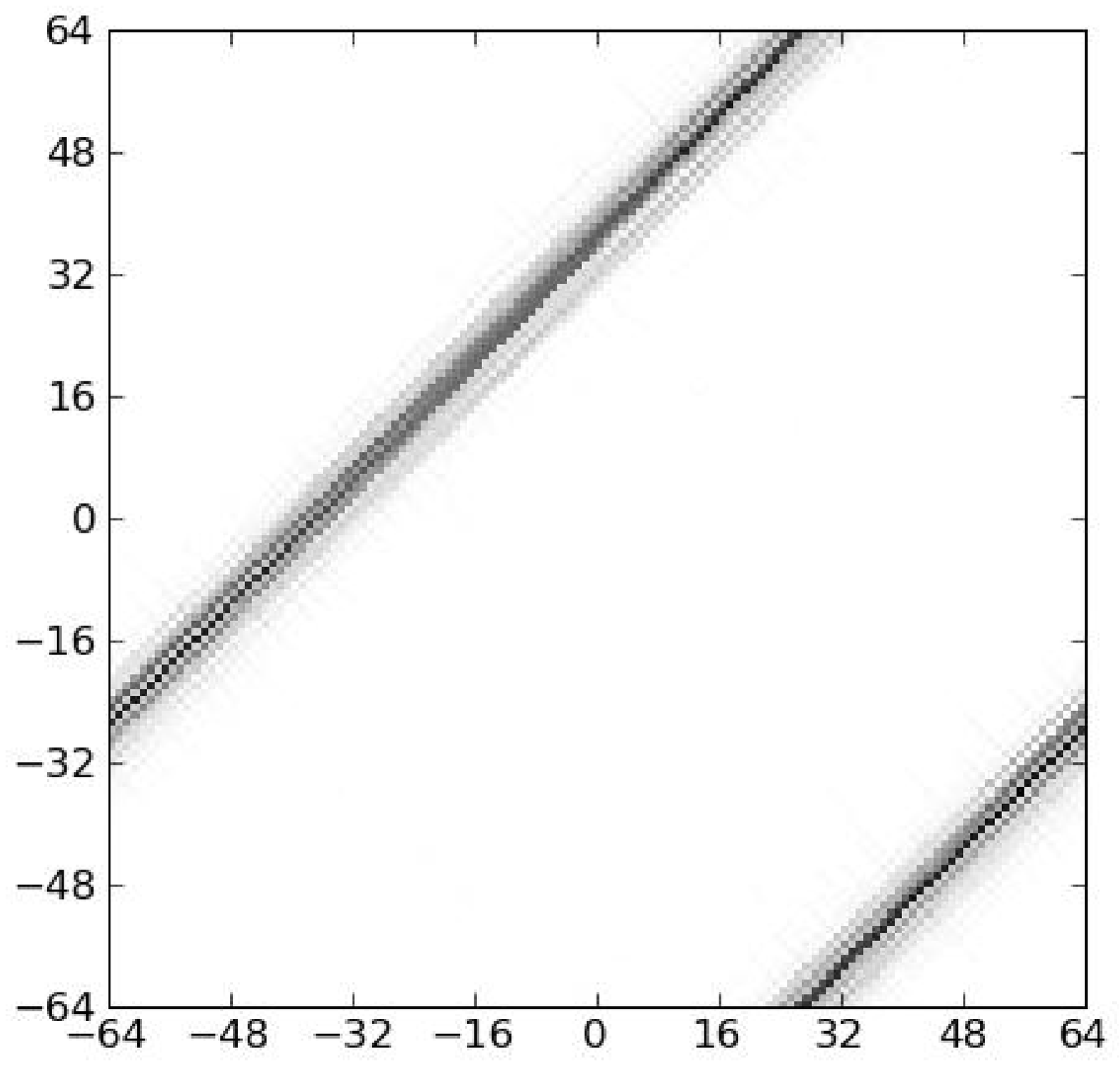}
\\
\vspace{2mm}\hspace{-57mm}
   \includegraphics[width=50mm]{ColorBarAct.eps}
  \caption{Maps of total cumulated plastic activity corresponding to
    increasing levels of deformation: $\varepsilon=0.25,1.0,4.0$. The
    initial threshold distribution is here shifted by
    $\varsigma_0=0.5$. The system size is $L=128$. A clear persistent
    shear band can be observed.}
  \label{persistent-localization}
\end{figure}

An implicit hypothesis performed in our model is that the statistical
distribution used to renew the local plastic threshold under shear ({\it i.e.}
after local slip) is the very same as the distribution of plastic thresholds in
the initial configuration. The evidence of this hypothesis may be questioned.
Indeed, various experimental and numerical results obtained in friction or in
shearing granular material or complex
fluids~\cite{Baumberger-AdvPhys06,Falk-PRL07,Behringer-GM10} seem to show an
effect of the preparation of the material upon its behavior under shear. One may
think for instance at the effect of density of granular material: a loose
(dense) packing tends to exhibit hardening (softening) while under shear their
density progressively evolves toward a ``critical'' value. In order to test the
effect of our hypothesis we give in the following a bias to the initial
thresholds distribution and try to test its consequences. Practically speaking
the initial thresholds are drawn from a uniform distribution in the range
$[\varsigma_0,1+\varsigma_0]$ while the (uniform) distribution of renewed
threshold remains unchanged in the range $[0,1]$. A positive (negative) value of
$\varsigma_0$ is expected to induce some softening (hardening) behavior since
all threshold values above unity (below zero) should eventually be replaced by
thresholds within the interval $[0,1]$. We show below results obtained with a
system of size $N=128$, a maximum slip increment $d=1$ and a bias
$\varsigma_0=0.5$. The effect appears to be rather spectacular since as shown in
Fig.~\ref{persistent-localization} a clear persistent localization is now
obtained.

Let us only note here that the way the plastic activity gets localized along a
band is somewhat reminiscent of the behavior of an earlier model proposed by
T\"or\"ok and Roux \cite{Torok-PRL00}. In this study, the authors made evidence
for a weak breaking of ergodicity which they relate to the progressive building
in the thresholds landscape of a valley (along the shear band) surrounded by
ridges elevating significantly above the base level. Plastic activity thus tends
to be confined in the valley and can no longer fully explore the disordered
landscape.

